\documentclass[journal]{IEEEtran}

\usepackage{amsmath}
\usepackage{amsfonts}
\usepackage{amssymb}
\usepackage{verbatim}
\usepackage{multirow}
\usepackage{graphicx,floatrow}
\usepackage{subfigure}
\usepackage{epstopdf}

\usepackage{algorithmic}
\usepackage[linesnumbered,lined,ruled]{algorithm2e}
\usepackage{stfloats}

\usepackage{color}
\usepackage{cite}
\usepackage{indentfirst}
\usepackage{array}
\usepackage{bm}
\usepackage{pstricks}
\usepackage{pst-node}
\usepackage{url}
\usepackage{booktabs}
\usepackage{threeparttable}
\usepackage{rotating}
\usepackage{enumerate}

\usepackage{color,xcolor}

\usepackage[switch,pagewise]{lineno}
\begin{document}

\title{Naive Gabor Networks for Hyperspectral Image Classification}

%
\author{Chenying~Liu, 
        Jun~Li, 
         Lin~He, 
                 Antonio Plaza, 
                 Shutao Li, 
               and  Bo Li
        \thanks{
This paper has been accepted by IEEE Transactions on Neural Networks and Learning Systems (IEEE TNNLS). This work was supported in part by National Science Foundation of China under Grants 61771496, 61571195, and 61901208, in part by National Key Research and Development Program of China under Grant 2017YFB0502900, in part by Guangdong Provincial Natural Science Foundation under Grants 2016A030313254, 2016A030313516, and 2017A030313382, in part by Natural Science Foundation of Jiangxi China under Grant 20192BAB217003. (\textit{Corresponding authors: Jun Li; Lin He}).

Chenying Liu and Jun Li are with the Guangdong Provincial Key Laboratory of Urbanization and Geo-simulation, School of Geography and Planning, Sun Yat-sen University, Guangzhou, 510275, China (\textit{e-mails: sysuliuchy@163.com; lijun48@mail.sysu.edu.cn}).

Lin He is with the School of Automation Science and Engineering, South China University of Technology, Guangzhou, 510640, China (\textit{e-mail: helin@scut.edu.cn}).

Shutao Li is with the College of Electrical and Information Engineering, Hunan University, Changsha 410082, China (\textit{e-mail: shutao\_li@hnu.edu.cn}).

Antonio Plaza is with the Hyperspectral Computing Laboratory, Department of Technology of Computers and Communications, Escuela Polit\'{e}cnica, University of Extremadura, C\'{a}ceres, E-10071, Spain (\textit{e-mail: aplaza@unex.es}).

Bo Li is with the Beijing Key Laboratory of Digital Media, School of Computer Science and Engineering, and the State Key Laboratory of Virtual Reality Technology and Systems, Beihang University, Beijing 100191, China (\textit{e-mail: boli@buaa.edu.cn}).

        }
        }


\maketitle

\begin{abstract}
Recently, many convolutional neural network (CNN) methods have been designed for hyperspectral image (HSI) classification since CNNs are able to produce good representations of data, which greatly benefits from a huge number of parameters. However, solving such a high-dimensional optimization problem often requires a large amount of training samples in order to avoid overfitting. Additionally, it is a typical non-convex problem affected by many local minima and flat regions. To address these problems, in this paper, we introduce naive Gabor Networks or Gabor-Nets which, for the first time in the literature, design and learn CNN kernels strictly in the form of Gabor filters, aiming to reduce the number of involved parameters and constrain the solution space, and hence improve the performances of CNNs. Specifically, we develop an innovative phase-induced Gabor kernel, which is trickily designed to perform the Gabor feature learning via a linear combination of local low-frequency and high-frequency components of data controlled by the kernel phase. With the phase-induced Gabor kernel, the proposed Gabor-Nets gains the ability to automatically adapt to the local harmonic characteristics of the HSI data and thus yields more representative harmonic features. Also, this kernel can fulfill the traditional complex-valued Gabor filtering in a real-valued manner, hence making Gabor-Nets easily perform in a usual CNN thread. We evaluated our newly developed Gabor-Nets on three well-known HSIs, suggesting that our proposed Gabor-Nets can significantly improve the performance of CNNs, particularly with a small training set.
\end{abstract}


\section{Introduction} \label{sec:intro}

Over the past two decades, hyperspectral imaging has witnessed a surge of interest for Earth Observations due to its capability to detect subtle spectral information using hundreds of continuous and narrow bands, thus making it promising for some applications such as classification \cite{Li201DL-HSI-ClassificationReview,He2018Review}. In the early stages of HSI classification, most techniques were devoted to analyzing the data exclusively in the spectral domain, disregarding the rich spatial-contextual information contained in the scene \cite{Fauvel2013Review}. Then, many approaches were developed to extract spectral-spatial features prior to classification to overcome this limitation, such as morphological profiles (MPs) \cite{Benediktsson2005EMP}, spatial-based filtering techniques \cite{He2017DLRGF,Jia2018GaborHSI}, etc., which generally adopts hand-crafted features following by a classifier with predefined hyperparameters.

Recently, inspired by the great success of deep learning methods \cite{Liu2019CNNSAR,Kim2019CNNBlind}, CNNs have emerged as a powerful tool for spectral and spatial HSI classification \cite{Paoletti2019Pyramidal,Hamouda2019AdaptiveSize}. Different from traditional methods, CNNs jointly learn the information for feature extraction and classification with a hierarchy of convolutions in a data-driven context, which is capable to capture features in different levels and generate more robust and expressive feature representations than the hand-crafted ones. Furthermore, the parameters can be optimized in accordance with data characteristics, leading to more effective models. However, CNN methods often require a large number of training samples in order to avoid overfitting, due to the huge number of free parameters involved. This is particularly challenging in the context of HSI classification, where manual annotation of samples is difficult, expensive and time-consuming. 
Moreover, solving the kernels of a CNN is a typical non-convex problem, affected by many local minima and flat regions \cite{Blum1993NPComplete}, which is usually addressed with a local search algorithm, such as the gradient descend algorithm under a random initialization scheme, making the kernels very likely converge to a bad/spurious local minimum  \cite{Sinom2018SupriousLocalMinima}.

To tackle these issues, a recent trend is to embed \textit{a priori} knowledge into deep  methods to refine model architectures. For example, Shamir \cite{Shamir2016GaussianInput} and Tian \cite{Tian2017GaussianInput} showed that the adoption of Gaussian assumptions on the input distribution can assist the successful training of neural networks. Chen \emph{et al.} \cite{Chen2018MMDP} overcame the contradiction between a small training size and a large parameter space through the integration of Bayesian modeling into neural networks. These previous works reveal that \emph{a priori} knowledge exhibits good potential in improving the reliability and generalization of deep models. More specifically, for CNNs some attempts have been made to reinforce model robustness via redesigning convolutional kernels using certain \emph{a priori} knowledge. For instance, circular harmonics are employed to equip CNNs with both translation and rotation equivariant kernels \cite{Worrall2017HNet}. However, the construction of such rotation-equivariant kernels is somewhat complicated, where each filter is a combination of a set of filter bases. Besides, the complex-valued convolution operations require a new CNN thread and increase the calculation burden in both the forward and backward propagations when using the same number of kernels.

\begin{figure}[tp]
  \centering
  \includegraphics[width=3.5in]{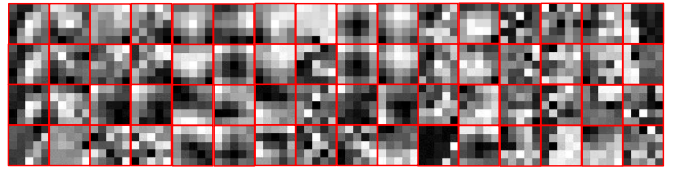}\\
  \caption{Filters extracted from the first convolutional layer of a well-trained CNN using 100 training samples per class for a hyperspectral image collected over Pavia University, Italy.}
  \label{fig:1stlayerKer}
\end{figure}

Apart from circular harmonics, Gabor filters offer another type of \emph{a priori} knowledge that can be used to reinforce convolutional kernels of CNNs. Gabor filters are able to achieve optimal joint time-frequency resolution from a signal processing perspective \cite{Gabor1946}, thus being appropriate for low-level and middle-level feature extractions (which are exactly the functions of the bottom layers of CNNs). Furthermore, researches have revealed that the shape of Gabor filters is similar to that of receptive fields of simple cells in the primary visual cortex \cite{Hubel1965CatCortex, Jones1987EvaluationCat, Pollen1983Bio, Alex2012AlexNet}, which means that using Gabor filters to extract low-level and middle-level features can be associated with a biological interpretation. In fact, as illustrated by Fig. \ref{fig:1stlayerKer}, many filters in CNNs (especially those in the first several layers), look very similar to Gabor filters. Inspired by these aspects, some attempts have been made to utilize the Gabor \textit{a priori} knowledge to reinforce CNN kernels, i.e. by replacing some regular kernels with fixed Gabor filters to reduce the number of parameters \cite{Jiang2018GCNNBinary, Calderon2003GCNNHandwritten, Sarwar2017GCNNEnergy}, initializing regular kernels in the form of Gabor filters \cite{Jiang2018GCNNBinary, Chang2014GCNNSpeech}, and modulating regular kernels with predefined Gabor filters \cite{Luan2018GCN}. Their good performance indicates a promising potential of Gabor filters in promoting the capacity of CNN models. However, traditional Gabor filtering is complex-valued, while CNNs are usually fulfilled with real-valued convolutions. Therefore, most Gabor-related CNNs only utilize the real parts of Gabor filters to form CNNs, which means that they only collect local low-frequency information in the data while disregarding (possibly useful) high-frequency information. To mitigate these issues, in analogy with some shallow-learning based Gabor approaches \cite{He2017DLRGF}, Jiang \textit{et al.} \cite{Jiang2018GCNNBinary} used the direct concatenation of the real and imaginary parts in CNNs. However, this approach is unable to tune the relationship between these two parts when extracting Gabor features. Most importantly, all these Gabor-related methods still manipulate hand-crafted Gabor filters, whose parameters are empirically set and remain unchanged during the CNN learning process. That is, the Gabor computation (although involved in these existing Gabor-related CNN models) does not play a significant role and, hence, is independent of the CNN learning. The remaining question is how to conveniently and jointly utilize the Gabor representation and the learning ability of CNNs so as to generate more effective features in a data-driven fashion.

In this work, we introduce a new concept of \textit{naive Gabor Networks} (or Gabor-Nets) for HSI classification where \textit{naive} refers to the fact that we straightforwardly replace regular convolutional kernels of CNNs with Gabor kernels, which is based on the following intuitions:
\begin{itemize}
\item First, Gabor filtering can be fulfilled with a linear convolution \cite{Arya2018ReviewGaborFuzzySVM}, which implies that Gabor filtering can be naturally extended to implement the basic convolution operations in CNNs.
\item Second, transforming the problem of solving CNN kernels to that of finding the optimal parameters of Gabor kernels tends to reduce the number of free parameters.
    If Gabor kernels, instead of regular CNN kernels, are used in a CNN, then the parameters to solve in each kernel will be transformed from the CNN kernel elements to Gabor parameters, such as the frequency magnitude, frequency direction and scale.
\item Although usually CNNs require real-valued computations (while Gabor filtering involves complex-valued computations related to the real and imaginary parts), there is a possibility to design flexible Gabor representations computed in a real-valued fashion without missing any information on the real and imaginary parts. This is because the local cosine harmonic and the local sinusoidal harmonic in a Gabor filter can be connected with each other by a phase offset term.
\end{itemize}
It is noteworthy that remotely sensed images are mainly composed of a series of geometrical and morphological features, i.e., low-level and middle-level features. Therefore, our networks (with a few Gabor convolutional layers) are expected to be able to extract representative features for HSI processing. To the best of our knowledge, this is the first attempt in the literature to both design and learn CNN convolutions strictly in the form of Gabor filters.

More specifically, the innovative contributions of our newly developed Gabor-Nets can be summarized as follows:
\begin{itemize}
\item Gabor-Nets operate in a twofold fashion. On the one hand, using Gabor filtering to perform convolutions in CNNs tends to reduce the number of the parameters to learn, thus requiring a smaller training set and achieving faster convergence during the optimization. On the other hand, the free parameters of Gabor filters can be automatically determined by the forward- and backward-propagations of CNNs.
\item Gabor-Nets are built on novel phase-induced Gabor kernels. The Gabor kernels, induced with a kernel phase term, exhibit two important properties. First, they have potential to adaptively collect both the local cosine and the local sinusoidal harmonic characteristics of the data. Second, the kernels can be used for real-valued convolutions. Thus, Gabor-Nets implemented with the new kernels are able to perform similarly to CNNs while generating more representative features.
\item Gabor-Nets adopt a well-designed initialization scheme. Specifically, the parameters used to construct usually-used hand-crafted Gabor filter banks are initialized based on the Gabor \textit{a priori} knowledge, while the kernel phases utilize a random initialization in order to increase the diversity of kernels. Such an initialization scheme can not only make Gabor-Nets inherit the advantages of traditional Gabor filters, but also equip Gabor-Nets with some superior properties such as the ability against the gradient vanishing problem often arising in CNNs.
\end{itemize}

The remaining of the paper is organized as follows. Section \ref{sec:relatedWorks} gives the general formulation of Gabor harmonics and simultaneously reviews some related works on Gabor filtering. Section \ref{sec:meth} introduces the proposed Gabor-Nets constructed with an innovative phase-induced Gabor kernel in detail. Section \ref{sec:exp} describes the experimental validation using three real hyperspectral datasets. Finally, section \ref{sec:con} concludes the paper with some remarks and hints at plausible future research lines.

\section{Related work} \label{sec:relatedWorks}

Let $(x, y)$ denote the space domain of an image. A {general} 2-D Gabor filter can be mathematically formulated by a Gaussian envelope modulated sinusoid harmonic, as follows:
\begin{equation}
\label{eq:2Dgabor:general2D}
\begin{aligned}
    \mathbf{G}(x,y) = & \frac{1}{2\pi\sigma_x\sigma_y}\exp{\bigg\{\!\!-\frac{1}{2}\Big(\frac{x_r^2}{\sigma_x^2}+\frac{y_r^2}{\sigma_y^2}\Big)\bigg\}} \\
    & \times\exp{\{j(x\omega_x+y\omega_y)\}},
\end{aligned}
\end{equation}
where $\sigma_x$ and $\sigma_y$ are the scales along the two spatial axes of the Gaussian envelope, $x_r=x\cos\phi+y\sin\phi$ and $y_r=-x\sin\phi+y\cos\phi$ are the rotated coordinates of $x$ and $y$ with a given angle $\phi$, $\omega_x=|\bm{\omega}|\cos\theta$ and $\omega_y=|\bm{\omega}|\sin\theta$ are the projections of a given angular frequency $\bm{\omega}$ onto  $x$ and $y$-directions, respectively, $\theta$ is the angle between $\bm{\omega}$ and the $x$-direction, $|\bm{\omega}|=(\omega_x^2+\omega_y^2)^\frac{1}{2}$ is the magnitude of $\bm{\omega}$ (hereinafter replaced by $\omega$) and $j$ is the imaginary unit. To simplify the gradient calculation of $\theta$, we utilize the rotation-invariant Gaussian envelope under unrotated coordinates with $\phi=0$ and $\sigma_x\!=\!\sigma_y\!=\!\sigma$. With Euler's relation, let $M=x\omega_x+y\omega_y$, we can rewrite the 2-D Gabor filter in the following complex form:
\begin{equation}
\label{eq:2Dgabor:euler}
\begin{aligned}
    \mathbf{G}(x,y) & = K \times\exp{\{jM\}}\\
    &= K\cos{M} +jK\sin{M}\\
    &= \Re\{\mathbf{G}(x,y)\} +j\Im\{\mathbf{G}(x,y)\},
\end{aligned}
\end{equation}
where $K =\frac{1}{2\pi\sigma^2}\exp{\{-\frac{x^2+y^2}{2\sigma^2}\}}$ is the rotation-invariant Gaussian envelope. Specifically, the local cosine harmonic $\Re\{\mathbf{G}(x,y)\}$ is associated with the local low-frequency component in the image, and the local sinusoidal harmonic $\Im\{\mathbf{G}(x,y)\}$ is connected to the local high-frequency component \cite{He2017DLRGF}, thus enabling Gabor filtering to access the local harmonic characteristics of the data.

In the following, we review some existing works relevant to the Gabor filtering.

\subsection{Traditional hand-Crafted Gabor filters} \label{sec:relatedWorks:GaborFilter}

From a signal processing perspective, Gabor harmonics maximize joint time/frequency or space/frequency resolutions \cite{Gabor1946}, making them ideal for computer vision tasks. Hand-crafted Gabor filters have achieved a great success in many applications, such as texture classification \cite{Idrissa2002GaborTexture}, face and facial expression recognition \cite{See2017GaborFace}, palmprint recognition \cite{Younesi2017GaborPalm}, edge detection \cite{Namuduri1992GaborEdge}, and several others \cite{Sun2005GaborVehicle}.
Regarding HSI data interpretation, Bau \textit{et al.} \cite{Bau2010RealGabor} used the real part of 3D Gabor filters to extract the energy features of regions for HSI classification, suggesting the effectiveness of Gabor filtering in feature extraction. He \textit{et al.} \cite{He2017DLRGF} proposed a novel discriminative low-rank Gabor filtering (DLRGF) method able to generate highly discriminative spectral-spatial features with high computational efficiency, thus greatly improving the performance of Gabor filtering on HSIs. Jia \textit{et al.} \cite{Jia2018GaborHSI} also achieved good classification results using the phase of complex-valued Gabor features. These hand-crafted Gabor features can be regarded as single-layer features extracted by Gabor filter banks. The involved parameters are empirically set following a "search strategy" where the orientations and spatial frequencies obey certain uniform distributions, aimed to cover as much optimal parameters as possible.

\subsection{Gabor-Related CNNs}  \label{sec:relatedWorks:GaborCNN}

\begin{table*}
  \scriptsize
  \begin{center}
  \renewcommand\arraystretch{1.0}
  \begin{tabular}{m{1.1cm}<{\centering}|m{2.5cm}<{\centering}||m{5.0cm}<{\centering}||m{2.5cm}<{\centering}||m{3.5cm}<{\centering}}
  \hline\hline
  \multicolumn{2}{c||}{Networks} & Functions of Gabor filters in CNNs  & \multicolumn{1}{c||}{\multirow{1}{*}{Type of Gabor filters}}  &  \multicolumn{1}{c}{\multirow{1}{*}{Setting of Gabor Parameters}}\\
  \hline\hline
  \multicolumn{1}{c|}{\multirow{4}{*}{Gabor-based}} & Using Gabor features & Feature extraction for inputs\cite{Hosseini2018InputGaborAge, Yao2016InputGaborOR, Lu2017InputGaborFace, Rizvi2016GCNNOR, Chen2017InputGaborHSI, Shi2018InputGaborShip} &  \multicolumn{1}{c||}{\multirow{4}{*}{Hand-crafted}}&  \multicolumn{1}{c}{\multirow{4}{*}{Fixed}}\\
  \cline{2-3}
  &                     & Fixed filters in shallow layers \cite{Jiang2018GCNNBinary, Calderon2003GCNNHandwritten, Sarwar2017GCNNEnergy}; &  & \\
  & Using Gabor filters & Initialization of kernels \cite{Chang2014GCNNSpeech};                  & & \\
  &                     & Modulation to kernels \cite{Luan2018GCN}       & & \\
  \hline
  \multicolumn{2}{c||}{Gabor-Nets (the proposed)} & Convolutional kernels  &  Learnable &  Tunable\\
  \hline\hline
  \end{tabular}
  \end{center}
  \caption{Comparison between the Gabor-based CNNs and the proposed Gabor-Nets.}
  \label{table:diff}
\end{table*}

Recently, some attempts have been made to incorporate Gabor harmonics into CNNs, in order to reduce the number of parameters and equip CNNs with orientation and frequency selectivity. The existing Gabor-related CNNs can be roughly categorized into two groups, i.e., those using Gabor features and those using Gabor filters. The former category uses Gabor features only as the inputs of networks; while in the latter category, predefined Gabor filters with fixed parameters are used in some convolutional layers.

Researches show that using hand-crafted Gabor features could help mitigate the negative effects introduced by a lack of training samples in CNNs. For example, Hosseini \emph{et al.} \cite{Hosseini2018InputGaborAge} utilized additional Gabor features as inputs for CNN-based age and gender classification, and obtained better results. Yao \emph{et al.} \cite{Yao2016InputGaborOR} achieved a higher recognition rate by using Gabor features to pre-train CNNs before fine-tuning. Similar works can be found in \cite{Lu2017InputGaborFace, Rizvi2016GCNNOR}. In the field of remote sensing, Chen \textit{et al.} \cite{Chen2017InputGaborHSI} fed the Gabor features extracted on the first several principal components into CNNs for HSI classification. Shi \emph{et al.} \cite{Shi2018InputGaborShip} complemented CNN features with Gabor features in ship classification. Their experimental results indicate that Gabor features are able to improve the performance of CNNs.

Another trend is to manipulate certain layers or kernels of CNNs with Gabor filters. For example, Jiang \emph{et al.} \cite{Jiang2018GCNNBinary} replaced the kernels in the first layer of a CNN with a bank of Gabor filters under predefined orientations and spatial frequencies. These first-layer Gabor filters can be fixed, as explained in \cite{Calderon2003GCNNHandwritten}, or be tuned at each kernel element, like \cite{Chang2014GCNNSpeech} where, in fact, Gabor filters were used for initialization purposes. Moreover, to reduce the training complexity, Sarwar \emph{et al.}, \cite{Sarwar2017GCNNEnergy} replaced some kernels in the intermediate layers with fixed Gabor filters and yielded better results. More recently, Luan \emph{et al.} \cite{Luan2018GCN} utilized Gabor filters to modulate regular convolutional kernels, thus making the network capable to capture more robust features with regards to orientation and scale changes. However, Gabor convolutional networks (GCNs) still learned the regular convolutional kernels, i.e., GCNs in fact utilized Gabor-like kernels.

These Gabor-related works, either using Gabor features or using Gabor filters, manipulate the hand-crafted Gabor filters without Gabor feature learning, which means that their parameters are empirically set (and remain unchanged) during the learning process. That is, in these existing Gabor-related CNNs, the Gabor computation does not play any relevant role in the CNN learning. In contrast, as illustrated in Table \ref{table:diff}, our proposed Gabor-Nets directly use Gabor kernels with free parameters as CNN kernels, and automatically determine the Gabor parameters with forward- and backward-propagations of CNNs in a data-driven fashion, thus being able to not only use the Gabor \textit{a priori} knowledge, but also to fulfill Gabor feature learning, therefore being adaptive to specific datasets and able to reduce the human supervision.

\section{Proposed Method}  \label{sec:meth}

In this section, we first introduce an innovative phase-induced Gabor kernel, followed by a discussion on its superior frequency properties. Then, we describe the proposed Gabor-Nets in detail.

\subsection{Phase-induced Gabor} \label{sec:meth:gkernel}

The real and imaginary parts of commonly-adopted Gabor filters are associated with the local low-frequency and high-frequency information in the data, respectively \cite{He2017DLRGF}. Some Gabor methods use only the real part to extract features, which obviously discards the possibility of exploiting local high-frequency information. In order to integrate the two components, other methods utilize the amplitude feature \cite{He2017DLRGF}
\begin{equation}
\label{eq:2Dgabor:amplitude}
\begin{aligned}
    \|\mathbf{G}(x,y)\| = \sqrt{\Re^2\{\mathbf{G}(x,y)\} +\Im^2\{\mathbf{G}(x,y)\}},
\end{aligned}
\end{equation}
the phase feature \cite{Jia2018GaborHSI}
\begin{equation}
\label{eq:2Dgabor:phase}
\begin{aligned}
    \sphericalangle\mathbf{G}(x,y) = \frac{\Im\{\mathbf{G}(x,y)\}}{\Re\{\mathbf{G}(x,y)\}},
\end{aligned}
\end{equation}
or the direct concatenation of the real and the imaginary parts \cite{Jiang2018GCNNBinary}. The latter case, though considering the real and the imaginary parts simultaneously (and hence synthesizing both the low-frequency and high-frequency information), is under a formulation where there is no parameter able to tune their relationship.

Additionally, as aforementioned, traditional Gabor filtering is connected to a complex-valued computation, whereas the standard CNNs are based on real-valued computations. Therefore, when applying Gabor kernels to CNNs, this difference has to be handled carefully.

\begin{figure}[htp]
\scriptsize
\centering
  \includegraphics[height=1.3in]{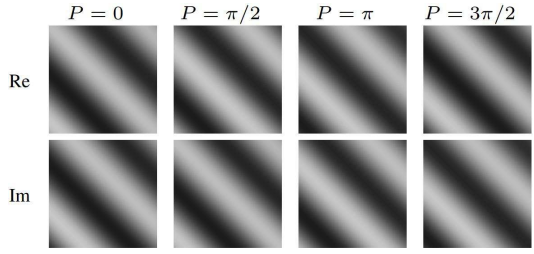}\\
  \caption{{Real and imaginary parts of complex-valued Gabor filters using $\theta=\pi/4$, $\omega=\pi/100$,  $\sigma=30$, where the imaginary part with a certain phase offset $P$ is just the corresponding real one with $(P-\pi/2)$.}}
\label{fig:gkernel:ReIm}
\end{figure}

In the following, we design an innovative phase-induced Gabor filter to deal with these problems. Let $P$ denote the phase offset of the sinusoidal harmonic. With $P$ added, the usually-used 2-D Gabor filtering formulated in (\ref{eq:2Dgabor:euler}) becomes
\begin{equation}
\label{eq:2Dgabor:general2DP}
\begin{aligned}
    \mathbf{G}_{P}(x,y) & = K \times\exp{\{j(M+P)\}}\\
    & = K\cos{(M+P)}+jK\sin{(M+P)}\\
    & = \Re\{\mathbf{G}_{P}(x,y)\} +j\Im\{\mathbf{G}_{P}(x,y)\},\\
\end{aligned}
\end{equation}
where we find,
\begin{equation}
\label{eq:equi:im}
\begin{aligned}
  K\sin(M+P) = K\cos\Big(M+(P-\frac{\pi}{2})\Big),
\end{aligned}
\end{equation}
that is,
\begin{equation}\label{eq:equi:relation}
  \Im\{\mathbf{G}_P(x,y)\} = \Re\{\mathbf{G}_{(P-\frac{\pi}{2})}(x,y)\}.
\end{equation}
As illustrated in (\ref{eq:equi:relation}) and Fig. \ref{fig:gkernel:ReIm}, there is a one-to-one correspondence in terms of $P$ between the real and the imaginary parts of $\mathbf{G}_{P}(x,y)$, i.e., the imaginary part with $P$ is the same as its real counterpart with a phase offset of $(P-\pi/2)$. Based on this observation, we develop a new Gabor filtering to serve as Gabor kernels in CNNs as follows,
\begin{equation}
\label{eq:gkernel:2D}
\begin{aligned}
    \mathbf{G}(x,y) = K\cos{(M+P)}.
\end{aligned}
\end{equation}
It can be observed that the Gabor filters with $P$=$0$ and $P$=$-\pi/2$ above are exactly the real and the imaginary parts of the traditional Gabor filters in (\ref{eq:2Dgabor:euler}), i.e., the low-frequency and the high-frequency components, respectively. This indicates that, with $P$ added, the Gabor filtering in (\ref{eq:gkernel:2D}), though only formulated with the cosine harmonic, can be equipped with different frequency properties as $P$ varies. Thus, we name this newly developed Gabor filtering as the phase-induced Gabor filtering. Obviously, this new Gabor filtering is fulfilled with a real-valued convolution, which means that, if we utilize such Gabor filters as Gabor kernels of a CNN, the traditional complex-valued Gabor computation can be avoided, hence allowing us to directly use Gabor kernels in a usual CNN thread. In this work, we refer to $P$ as the kernel phase of Gabor kernels.

For clarity, hereinafter we utilize $\mathbf{G}'(x,y)$ to represent the Gabor filters without a phase offset term in (\ref{eq:2Dgabor:general2D})-(\ref{eq:2Dgabor:phase}), with their real and the imaginary parts denoted as $\Re\{\mathbf{G}'(x,y)\}$ and $\Im\{\mathbf{G}'(x,y)\}$, respectively.

\subsection{Adaptive Frequency Response Property} \label{sec:meth:properties}

To obtain a deeper inspection of our innovative phase-induced Gabor kernel, we decompose (\ref{eq:gkernel:2D}) using the trigonometric formula as follows:
\begin{equation}
\label{eq:gkernel:sepP}
\begin{aligned}
    \mathbf{G}(x,y) & = \cos{P}\cdot K\cos{M}-\sin{P}\cdot K\sin{M}\\
    & = \cos{P}\cdot \Re\{\mathbf{G}'(x,y)\}-\sin{P}\cdot \Im\{\mathbf{G}'(x,y)\}.
\end{aligned}
\end{equation}
As can be observed, $\mathbf{G}(x,y)$ is actually a linear combination of $\Re\{\mathbf{G}'(x,y)\}$ and $\Im\{\mathbf{G}'(x,y)\}$ in (\ref{eq:2Dgabor:euler}), which involves the weights $\cos P$ and $\sin P$ controlled by the kernel phase $P$. Utilizing the forward- and backward- propagations of CNNs, the free parameter $P$ can be tuned following the data. Recall that $\Re\{\mathbf{G}'(x,y)\}$ and $\Im\{\mathbf{G}'(x,y)\}$ are associated with the local low-frequency and high-frequency components, respectively \cite{He2017DLRGF}. Thus, by involving the kernel phase $P$, we can develop the CNN constructed by Gabor kernels, which is able to adaptively process both the low-frequency and high-frequency characteristics of the data.

\begin{figure}[htp]
\scriptsize
  \centering
  \begin{tabular}{p{3.7cm}<{\centering}p{3.7cm}<{\centering}}
  \includegraphics[height=1.2in]{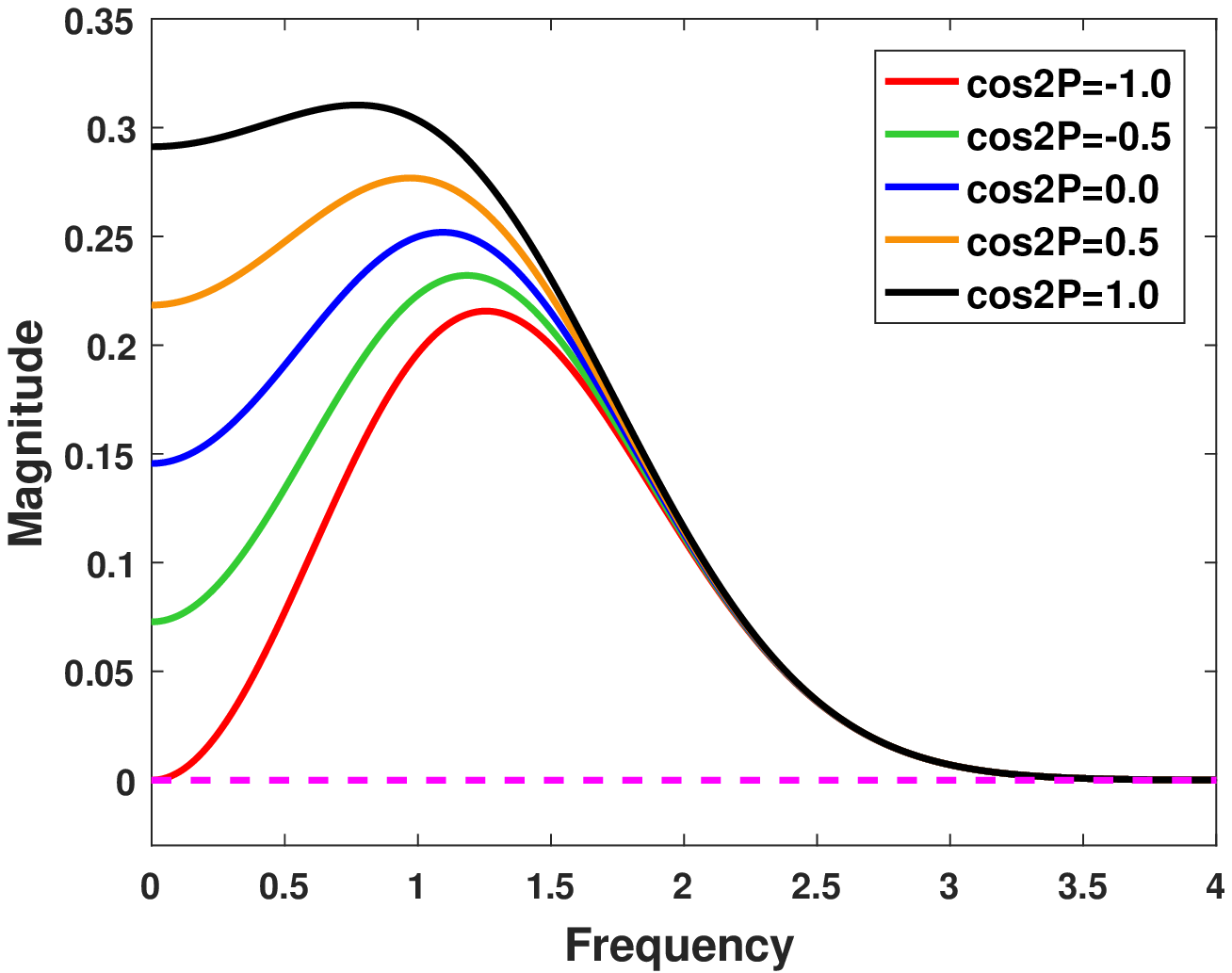} &
  \includegraphics[height=1.2in]{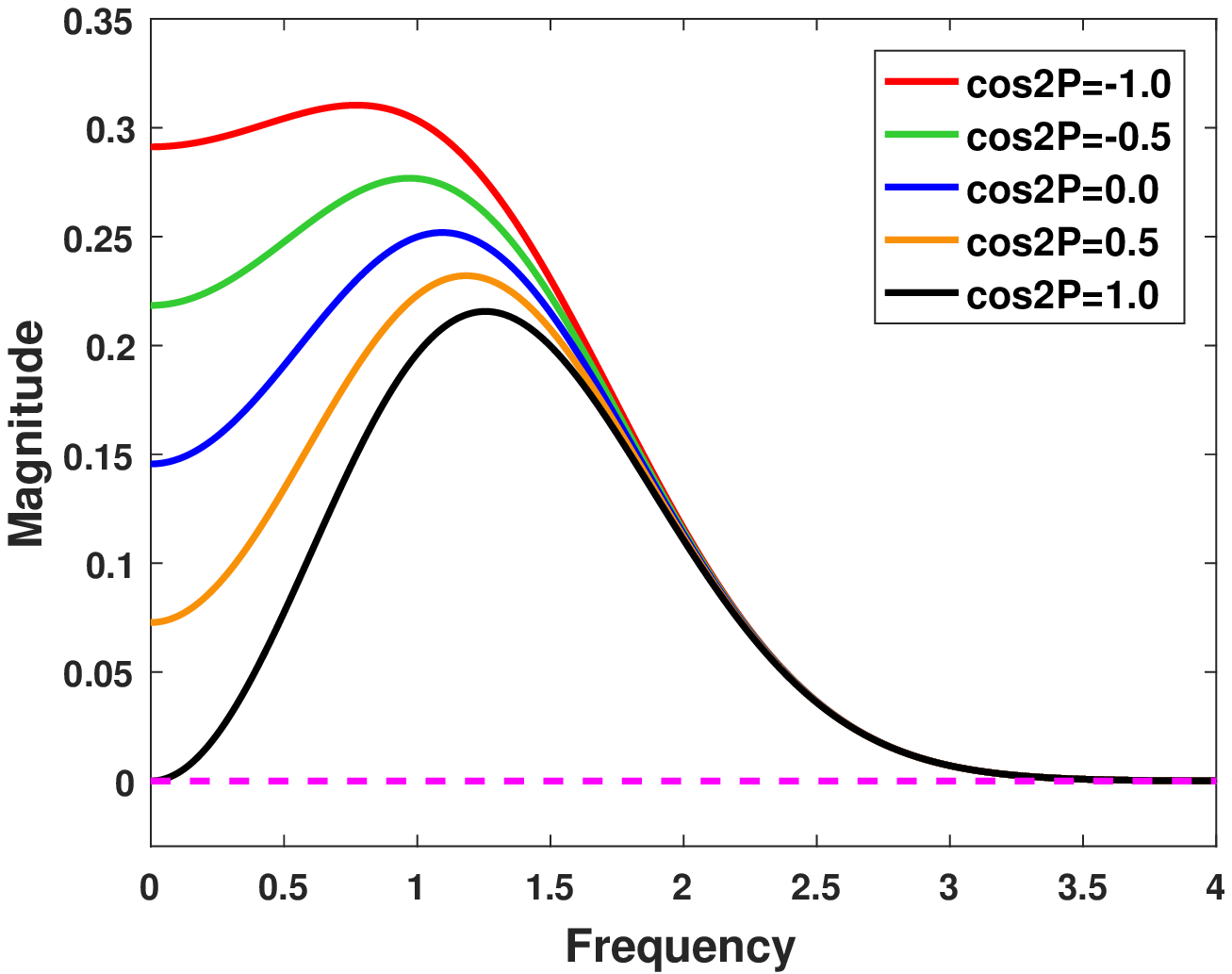} \\
  (a) & (b)
  \end{tabular}
  \caption{Squared frequency magnitudes of Gaussian enveloped (a) cosine and (b) sinusoidal harmonics with varying values of $P$.}
\label{fig:phase:mag}
\end{figure}

Reconsidering the decomposition of (\ref{eq:gkernel:2D}), it can be found that the cosine harmonic is formed by the coupling of $x$ and $y$. If we decouple $x$ and $y$ via the trigonometric formula and separate the Gaussian envelope $K$ along the $x$ and $y$-directions, respectively, (\ref{eq:gkernel:2D}) turns to \footnote{It is identical to allocate the phase offset $P$ with $x$ or $y$.}:
\begin{equation}
\label{eq:gkernel:decomp2}
\begin{aligned}
    \mathbf{G}=g_{c,p}^{(x)} \cdot g_{c}^{(y)}-g_{s,p}^{(x)} \cdot g_{s}^{(y)},
\end{aligned}
\end{equation}
where
\begin{equation}
\label{eq:gkernel:xpcos}
\begin{aligned}
    g_{c,p}^{(x)}=\frac{1}{\sqrt{2\pi}\sigma}\exp{\Big(-\frac{x^2}{2\sigma^2}\Big)}\cos{(x\omega_x+P)},
\end{aligned}
\end{equation}
\begin{equation}
\label{eq:gkernel:ycos}
\begin{aligned}
    \hspace{-0.7cm}g_{c}^{(y)}=\frac{1}{\sqrt{2\pi}\sigma}\exp{\Big(-\frac{y^2}{2\sigma^2}\Big)}\cos{(y\omega_y)},
\end{aligned}
\end{equation}
\begin{equation}
\label{eq:gkernel:xpsin}
\begin{aligned}
    g_{s,p}^{(x)}=\frac{1}{\sqrt{2\pi}\sigma}\exp{\Big(-\frac{x^2}{2\sigma^2}\Big)}\sin{(x\omega_x+P)},
\end{aligned}
\end{equation}
and
\begin{equation}
\label{eq:gkernel:ysin}
\begin{aligned}
    \hspace{-0.7cm}g_{s}^{(y)}=\frac{1}{\sqrt{2\pi}\sigma}\exp{\Big(-\frac{y^2}{2\sigma^2}\Big)}\sin{(y\omega_y)}.
\end{aligned}
\end{equation}
As proven in \cite{He2017DLRGF}, $g_{c}^{(y)}$ and $g_{s}^{(y)}$, without the kernel phase $P$, are low-frequency pass and low-frequency resistant filters, respectively. Regarding the other two components, i.e., $g_{c,p}^{(x)}$ and $g_{s,p}^{(x)}$, given $\omega_0$ of $\omega_x$, their mathematical representations in the frequency domain are calculated as follows,
\begin{equation}
\label{eq:FT:cos}
\begin{aligned}
    \widehat{g}_{c,p}(\omega) = \frac{1}{2}(A+B)\cos{P} + \frac{1}{2}j(A-B)\sin{P},
\end{aligned}
\end{equation}
 and
\begin{equation}
\label{eq:FT:sin}
\begin{aligned}
    \widehat{g}_{s,p}(\omega) = \frac{1}{2}(A+B)\sin{P} - \frac{1}{2}j(A-B)\cos{P},
\end{aligned}
\end{equation}
where $A\!=\!\exp\!{\big(\!-\!\frac{\sigma^2(\omega-\omega_0)^2}{2}\!\big)}$, and $B\!=\!\exp\!{\big(\!-\!\frac{\sigma^2(\omega+\omega_0)^2}{2}\!\big)}$. Then their corresponding squared magnitudes of frequency can be obtained as follows,
\begin{equation}
\label{eq:FT:cos:mag}
\begin{aligned}
    |\widehat{g}_{c,p}(\omega)|^2 & = \frac{1}{4}\!\exp{\big(\!-\!{\sigma^2(\omega\!-\!\omega_0)^2}\big)} \!+\! \frac{1}{4}\!\exp{\big(\!-\!{\sigma^2(\omega\!+\!\omega_0)^2}\big)}\\
    &\;\;\;\;+ \frac{1}{2}\cos{(2P)}\exp{\big(-\sigma^2(\omega^2+\omega_0^2)\big)},
\end{aligned}
\end{equation}
\begin{equation}
\label{eq:FT:sin:mag}
\begin{aligned}
    |\widehat{g}_{s,p}(\omega)|^2 & = \frac{1}{4}\!\exp{\big(\!-\!{\sigma^2(\omega\!-\!\omega_0)^2}\big)}\!+\!\frac{1}{4}\!\exp{\big(\!-\!{\sigma^2(\omega\!+\!\omega_0)^2}\big)}\\
    &\;\;\;\;- \frac{1}{2}\cos{(2P)}\exp{\big(-\sigma^2(\omega^2+\omega_0^2)\big)}.
\end{aligned}
\end{equation}
If we set $\omega$ in (\ref{eq:FT:cos:mag}) and (\ref{eq:FT:sin:mag}) to be zero, we have
\begin{equation}
\label{eq:FT:cos:mag0}
\begin{aligned}
    |\widehat{g}_{c,p}(0)|^2
    & = \frac{1}{2}[1+\cos(2P)]\exp{\big(-\sigma^2\omega_0^2\big)},
\end{aligned}
\end{equation}
\begin{equation}
\label{eq:FT:sin:mag0}
\begin{aligned}
    |\widehat{g}_{s,p}(0)|^2
    & = \frac{1}{2}[1-\cos(2P)]\exp{\big(-\sigma^2\omega_0^2\big)}.
\end{aligned}
\end{equation}
As shown in (\ref{eq:FT:cos:mag0}) and (\ref{eq:FT:sin:mag0}), when $\cos(2P)$ is approaching $-1$, i.e., $P$ is approaching $\pi/2$, $|\widehat{g}_{c,p}(0)|^2$ and $|\widehat{g}_{s,p}(0)|^2$ are decreasing to and increasing away from $0$, respectively (which implies that the low-frequency resistance of $g_{c,p}^{(x)}$ is being enforced, while $g_{s,p}^{(x)}$ behaves more like a low-pass filter). The situation is opposite when $\cos(2P)$ is approaching $1$, i.e., $P$ is approaching 0.
Fig. \ref{fig:phase:mag} shows the appearance of the squared magnitudes in the frequency domain with varying values of $P$. Clearly, {the frequency response characteristics} of $g_{c,p}^{(x)}$ and $g_{s,p}^{(x)}$ significantly change as $P$ varies.

For comparison, we employ the same strategy to decompose (\ref{eq:2Dgabor:euler}) as follows,
\begin{equation}
\label{eq:2Dgabor:noPdecompRe}
\begin{aligned}
    \Re\{\mathbf{G}^{'}(x,y)\} = g_{c}^{(x)}\!\cdot\!g_{c}^{(y)}-g_{s}^{(x)}\!\cdot\!g_{s}^{(y)},
\end{aligned}
\end{equation}
\begin{equation}
\label{eq:2Dgabor:noPdecompIm}
\begin{aligned}
    \Im\{\mathbf{G}^{'}(x,y)\} = g_{s}^{(x)}\!\cdot\!g_{c}^{(y)}+g_{c}^{(x)}\!\cdot\!g_{s}^{(y)},
\end{aligned}
\end{equation}
where
\begin{equation}
\label{eq:gkernel:xcos}
\begin{aligned}
    g_{c}^{(x)}=\frac{1}{\sqrt{2\pi}\sigma}\exp{\Big(-\frac{x^2}{2\sigma^2}\Big)}\cos{(x\omega_x)},
\end{aligned}
\end{equation}
\begin{equation}
\label{eq:gkernel:xsin}
\begin{aligned}
    g_{s}^{(x)}=\frac{1}{\sqrt{2\pi}\sigma}\exp{\Big(-\frac{x^2}{2\sigma^2}\Big)}\sin{(x\omega_x)}.
\end{aligned}
\end{equation}
It can be observed that, (\ref{eq:2Dgabor:euler}) is composed of a series of components which all have an explicit frequency nature. Accordingly, if we utilize the Gabor filtering without $P$ in (\ref{eq:2Dgabor:euler}) as Gabor kernels, the fundamental properties of kernels, i.e., the frequency response characteristics, can hardly be changed though other parameters are adaptively tuned in the learning process.

\begin{figure*}[htp]
\scriptsize
  \centering
  \includegraphics[width=0.90\linewidth]{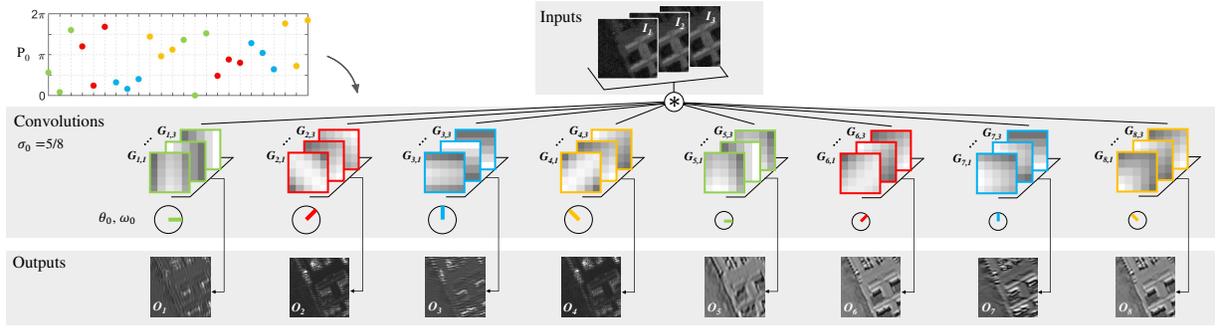}
  \caption{Forward process of one convolutional layer in Gabor-Nets, with the number of  input features set to $N_i=3$ and the kernel size set to $k=5$. The kernels are constructed using four $\theta_0$s and two $\omega_0$s, \emph{i.e.}, $N_t=4$ and $N_m=2$, where the ones marked by circles with the same color and the same size are initialized with the same $\theta_0$ and $\omega_0$, respectively. Therefore, there are 8 output features in total, \emph{i.e.}, $N_0=4\times 2=8$. Furthermore,  $\sigma$s are all initialized with $k/8=5/8$, and  $P$s are randomly initialized within the range $[0,2\pi)$.}
  \label{fig:GaborNet}
\end{figure*}

To conclude the above two parts, the roles of the kernel phase $P$, which is crucial in our phase-induced Gabor kernel and therefore the newly developed Gabor-Nets, can be summarized as follows:
\begin{itemize}
\item The kernel phase $P$ endows Gabor kernels with the ability to adaptively collect both the local cosine and the local sinusoidal harmonic characteristics of the data, via adjusting the their linear combination.
\item With the kernel phase $P$, the traditional complex-valued Gabor kernel can be fulfilled in a real-valued manner, therefore making it possible to directly (and conveniently) utilize our phase-induced Gabor kernel to construct a real-valued CNN.
\end{itemize}

\subsection{Gabor-Nets} \label{sec:meth:gabornet}

The proposed Gabor-Nets directly use phase-induced Gabor kernels to fulfill CNN convolutions.
Then, the parameters to solve in each convolutional kernel are transformed from the kernel elements \emph{per se} to the Gabor parameters of a phase-induced Gabor kernel: $\{\theta, \omega, \sigma, P\}$, i.e., the angle between the angular frequency and the $x$-direction $\theta$, the magnitude of the angular frequency $\omega$, the scale $\sigma$, and the kernel phase $P$. Let $k$ denote the kernel size. A phase-induced Gabor kernel has only four parameters to learn no matter how $k$ varies, whereas a regular kernel has $k^2$ elements to solve. In the situation with a smallest kernel size (the $1\times1$ kernels are not considered in this work), i.e., when $k=3$, the numbers of free parameters of a Gabor kernel and a regular kernel are 4 and 9, respectively. With the increase of $k$, the difference between those parameter numbers increases.
For simplicity, we utilize $\mathbf{G}$ in place of $\mathbf{G}(x,y)$, and $\{\theta_0, \omega_0, \sigma_0, P_0\}$ to represent the initializations of corresponding Gabor parameters hereinafter. As illustrated in Fig. \ref{fig:GaborNet}, the number of output features in the $l$th convolutional layer of Gabor-Nets is determined as $N_o=N_t \times N_m$, where $N_t$ and $N_m$ are the predefined numbers of $\theta_0$s and $\omega_0$s, respectively.
Then, with $N_i$ input features, the kernels in the $l$th layer are defined as
\begin{equation}
\label{eq:GaborNet:kernels}
\begin{aligned}
    \mathbf{G}^{(l)}=\{ \mathbf{G}_{1}^{(l)},\mathbf{G}_{2}^{(l)},\cdots,\mathbf{G}_{N_o}^{(l)}\},
\end{aligned}
\end{equation}
where $\mathbf{G}_{o}^{(l)}=\{\mathbf{G}_{o,1}^{(l)},\cdots,\mathbf{G}_{o,N_i}^{(l)} \}, o = 1,2,\cdots,N_o$ is the $o$th kernel, i.e., a set of $N_i$ Gabor filters corresponding to $N_i$ input features used to generate the $o$th output feature. Within a kernel, the $N_i$  filters are initialized with the same $\theta_0$ and $\omega_0$, and then are fine-tuned in a data-driven context during the training process. As a result, we can obtain the output features as follows,
\begin{equation}
\label{eq:GaborNet:output}
\begin{aligned}
    \mathbf{O}^{(l)}=\{\mathbf{O}_{1}^{(l)},\mathbf{O}_{2}^{(l)},\cdots,\mathbf{O}_{N_o}^{(l)} \},
\end{aligned}
\end{equation}
where
$\mathbf{O}_{o}^{(l)}=\sum_i^{N_i} \mathbf{I}_i^{(l)} \ast \mathbf{G}_{o,i}^{(l)}$ for $o=1,2,\cdots,N_o$, and $\mathbf{I}^{(l)}=\{\mathbf{I}_1^{(l)},\cdots,\mathbf{I}_{N_i}^{(l)}\}$ are the input features of the $l$th layer. For the first layer, $\mathbf{I}^{(l)}$ are the initial input features of the network, otherwise $\mathbf{I}^{(l)}=\mathbf{O}^{(l-1)}$.

Notice that the key difference between the proposed Gabor-Nets and regular CNNs is the designed form of convolutional kernels. Therefore, it is very easy to incorporate other CNN elements or tricks into Gabor-Nets, such as pooling, batch normalization, activation functions, etc.

\subsubsection{Initialization of Gabor kernels}

In order to guarantee the effectiveness of Gabor kernels, we provide a generally reliable initialization scheme for Gabor-Nets. First, following the "search strategy" used for the settings of hand-crafted Gabor filter banks, the $\theta_0$s are predefined as an evenly spaced sequence of $[0,\pi)$ based on $N_t$, and the $\omega_0$s are set as a geometric sequence with an initial value of $(\pi/2)$ and a geometric progression of $(1/2)$. For example, as shown in Fig. \ref{fig:GaborNet}, to construct a Gabor convolutional layer using $N_t=4$ and $N_m=2$, we set $\theta_0$s to be 0, $(\pi/4)$, $(\pi/2)$, $(3\pi/4)$, and $\omega_0$s to be $(\pi/2)$, $(\pi/4)$, respectively. On the one hand, the "search strategy" has been proven effective in traditional hand-crafted Gabor feature extraction by covering as many orientations and frequencies as possible. Then in Gabor-Nets, although each kernel is initially specific to one orientation and one frequency, different orientations and frequencies couple with each other as the layer goes deeper. On the other hand, such initializations are in accordance with the common observation that an HSI contains the information in many directions, while the discriminative information tends to appear on low frequencies \cite{He2017DLRGF}.
The initialization of $\sigma$s is relatively empirical among the four parameters. As stated above, $\sigma$ controls the localization scale of the filter. In hand-crafted Gabor filter design, $\sigma$ is always set to be one quarter of the kernel size. Taken into consideration the fact that CNNs generate features via multi-layer convolutions, we initialize $\sigma$s to be one eight of the kernel size in our work. Regarding the kernel phase $P$, we adopt a random initialization of $P$ in order to increase the diversity of Gabor kernels, aimed at promoting the robustness of Gabor-Nets. As indicated in (\ref{eq:gkernel:sepP}), $P$ dominates the harmonic characteristics of Gabor kernels via $\cos{P}$ and $\sin{P}$. Therefore, we randomly initialize $P_0$s within $[0,2\pi)$, i.e., both $\sin{P_0}$ and $\cos{P_0}$ within $[-1,1]$ in each layer.

\subsubsection{Updating of Gabor kernels}
In the back-propagation stage of Gabor-Nets, we update the convolutional kernels as a whole by solving the aforementioned Gabor parameters, the gradients of which are aggregated from all the elements of the kernel as follows:
\begin{equation}
\label{eq:GaborNet:error}
\begin{aligned}
    &\delta_\tau = \frac{\partial L}{\partial \tau}=\sum_{x,y}\mathbf{\delta}_{\mathbf{G}}\circ \frac{\partial \mathbf{G}}{\partial \tau},\\
    &\tau\leftarrow\tau-\delta_\tau, \;\;\;\;\text{for}\; \tau = \{\theta, \omega\,\sigma, P\},
\end{aligned}
\end{equation}
where $\mathbf{\delta}_{\mathbf{G}}$ is {the gradient of the training loss $L$ w.r.t. $\mathbf{G}$}, $\circ$ is the Hadamard product, and
\begin{equation}
\label{eq:GaborNet:PGrad}
\begin{aligned}
    \hspace{-0.3cm}\frac{\partial \mathbf{G}}{\partial P}&= -\frac{1}{2\pi\sigma^2}\exp{\big(-\frac{x^2+y^2}{2\sigma^2}\big)}\sin{(x\omega_x+y\omega_y+P)} \\ &=-K\sin{(x\omega_x+y\omega_y+P)},
\end{aligned}
\end{equation}
\begin{equation}
\label{eq:GaborNet:thGrad}
\begin{aligned}
    \hspace{-0.3cm}\frac{\partial \mathbf{G}}{\partial \theta}
     &= -\frac{1}{2\pi\sigma^2}\exp{\big(-\frac{x^2+y^2}{2\sigma^2}\big)}\sin{(x\omega_x+y\omega_y+P)}\\
     &\hspace{0.5cm} \cdot (-x\omega\sin\theta+y\omega\cos\theta)\\
     &= \frac{\partial \mathbf{G}}{\partial P}\circ(-x\omega_y+y\omega_x),
\end{aligned}
\end{equation}
\begin{equation}
\label{eq:GaborNet:omgGrad}
\begin{aligned}
    \hspace{-0.3cm}\frac{\partial \mathbf{G}}{\partial \omega}
     &= -\frac{1}{2\pi\sigma^2}\exp{\big(-\frac{x^2+y^2}{2\sigma^2}\big)}\sin{(x\omega_x+y\omega_y+P)}\\
     &\hspace{0.5cm}\cdot(x\cos\theta+y\sin\theta)\\
     &= \frac{\partial \mathbf{G}}{\partial P}\circ(x\cos\theta+y\sin\theta),
\end{aligned}
\end{equation}
\begin{equation}
\label{eq:GaborNet:sigGrad}
\begin{aligned}
\hspace{-0.3cm}\frac{\partial \mathbf{G}}{\partial \sigma}&= -\frac{1}{\pi\sigma^3}\cdot\exp{\big(-\frac{x^2+y^2}{2\sigma^2}\big)}\cos{(x\omega_x+y\omega_y+P)} \\
    &\hspace{0.8cm}+\frac{1}{2\pi\sigma^2}\cdot\frac{x^2+y^2}{\sigma^3}\exp{\big(-\frac{x^2+y^2}{2\sigma^2}\big)}\\
    &\hspace{1.5cm}\cdot\cos{(x\omega_x+y\omega_y+P)}\\
    &=(\frac{x^2+y^2}{\sigma^3}-\frac{2}{\sigma})\cdot \frac{1}{2\pi\sigma^2}\exp{\big(-\frac{x^2+y^2}{2\sigma^2}\big)}\\
    &\hspace{0.5cm}\cdot\cos{(x\omega_x+y\omega_y+P)}\\
    &=\mathbf{G}\circ(\frac{x^2+y^2}{\sigma^3}-\frac{2}{\sigma}).
\end{aligned}
\end{equation}

\section{Experiments}  \label{sec:exp}

In this section, we first present the experimental settings for the sake of reproduction, following which the proposed Gabor-Nets are evaluated using three real hyperspectral datasets, i.e. the Pavia University scene, the Indian Pines scene, and the Houston scene\footnote{The Pavia University scene and the Indian Pines scene can be downloaded from \url{https://www.sipeo.bgu.tum.de/downloads}. The Houston scene can be downloaded from \url{https://www.grss-ieee.org/community/technical-committees/data-fusion/2013-ieee-grss-data-fusion-contest/}}. After that, we investigate some properties of Gabor-Nets via relevant experiments.

\subsection{Experimental settings} \label{sec:exp:settings}

\begin{figure}[thp]
\scriptsize
  \centering
  \includegraphics[width=1.8in]{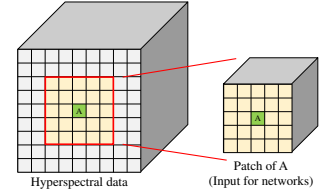}\\
  \caption{Graphical illustration of the patch generation. Take the patch size of 5 for example.}
  \label{fig:patchGene}
\end{figure}

To conduct the pixel-wise HSI classification with CNNs, patch generation is a widely used strategy in the preprocessing stage to prepare the inputs of networks \cite{Plaza20183DCNN,Lee2017DCCNN,Liu2018SCNN}. Let $S_p$ denote the patch size. The input patch is defined as $S_p\times S_p$ neighboring pixels centered on the given pixel. As shown in Fig. \ref{fig:patchGene}, taking $S_p$ of 5 for example, the patch of the given pixel A is the surrounding square area, each side of which is 5 pixels long (the red box in Fig. \ref{fig:patchGene}). Accordingly, the label/output of this patch is that of A.

\begin{figure}[thp]
\scriptsize
  \centering
  \includegraphics[height=1.85in]{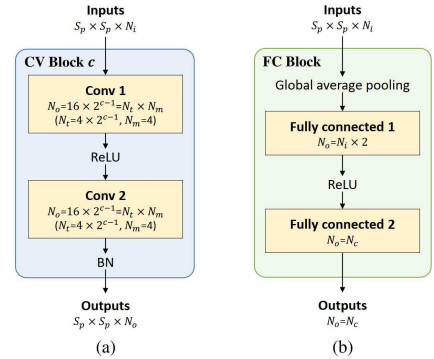}\\
  \caption{(a) Unit convolutional block (CV Block) and (b) fully connected block (FC Block) used for HSI classification network construction, where $S_p$ is the patch size, $c$ is the index of CV Blocks, $N_i$ and is the number of inputs, $N_t$ and $N_m$ are the numbers of $\theta_0$s and $\omega_0$s of Gabor convolutional layers, $N_o$ is the number of outputs for the layers/blocks.}
  \label{fig:para:struct}
\end{figure}

\begin{table}[ht]
\scriptsize
\begin{center}
\begin{tabular}{c|c|c}
\hline\hline
$\sharp$para      &  Regular CNNs  &  Gabor-Nets \\
\hline\hline
Conv1 & $\hspace{0.2cm}k^2\times N_i \times N_o + N_o$ & $\hspace{0.2cm}4\times N_i \times N_o + N_o$  \\
Conv2 & $\hspace{0.2cm}k^2\times N_o \times N_o$       & $\hspace{0.2cm}4\times N_o \times N_o$  \\
BN    & $2\times N_o$                                  & $2\times N_o$ \\
\hline
Total & $k^2(N_i+N_o)N_o+3N_o$   & $4(N_i+N_o)N_o+3N_o$\\
\hline\hline
\end{tabular}
\end{center}
\caption{Numbers of parameters used in a CV Block of the regular CNNs and Gabor-Nets, respectively, where $k$ is the kernel size.}
\label{table:cvbpara}
\end{table}

Regarding the architecture of networks, we utilized the unit convolutional block (CV Block) and the fully connected block (FC Block) illustrated in Fig. \ref{fig:para:struct} to construct the basic network architectures for regular CNNs and Gabor-Nets used in our experiments. Notice that regular CNNs and Gabor-Nets shared the same architectures, yet utilized different types of convolutional kernels. The former used regular kernels, while the latter used the proposed Gabor kernels. As shown in Fig. \ref{fig:para:struct}, each CV Block contains two convolutional (Conv) layers, one rectified linear unit (ReLU) nonlinearity layer, and one Batch Normalization (BN) layer. The CV Block is designed in accordance to \cite{Worrall2017HNet}. We utilized 16 kernels in each convolutional layers of the first CV Block. Each additional CV Blocks doubled the kernel number. For the initialization of Gabor-Nets, the number of $\theta_0$s of the first CV Block was set to be 4, and then doubled as more CV Blocks were added. The number of $\omega_0$s remained to be 4 in all the CV Blocks. The input number was the number of bands for the first CV block, otherwise equalled the output number of the last CV Block. No pooling layers were utilized in CV Blocks, since the patch size was relatively small in our experiments. As reported in Table \ref{table:cvbpara}, for each CV Block, regular CNNs contain $(k^2-4)(N_i+N_o)N_o$ more parameters than Gabor-Nets. The difference becomes larger as $k$, $N_i$ and $N_o$ increase.
On top of CV Blocks is one FC Block with two fully connected layers, one global average pooling layer, and one ReLU layer. In the FC Block, the global average pooling layer is first utilized to rearrange $N_i$ input feature maps into a vector of $N_i$ elements, in order to reduce the number of parameters of fully connected layers. The output number of the first fully connected layer is twice its input number, while that of the second fully connected layer equals the number of predefined classes for classification purposes. The FC Block is completely the same for Gabor-Nets and regular CNNs since it contains no Conv layers. The number of parameters in an FC Block is $(N_i\cdot 2N_i+2N_i)+(2N_i\cdot N_c+N_c)=2{N_i}^2+2N_i+2{N_i}{N_c}+N_c$.
In our experiments, We used the cross-entropy loss and Adam optimizer. The learning rate was initially set to be 0.0076, decaying automatically at a rate of 0.995 at each epoch. The total number of epochs is 300.

Except for the regular CNNs, we also considered some other state-of-the-art deep learning based HSI classification methods for comparison. The first one used the hand-crafted Gabor features as the inputs of the CNNs (Gabor as inputs) \cite{Chen2017InputGaborHSI}, where the Gabor features were generated from the first several principal components of HSI data. The hand-crafted parameters were set in accordance to the initializations of the first convolutional layer of Gabor-Nets. The second one is the deep contextual CNN (DC-CNN) \cite{Lee2017DCCNN}, leveraging the residual learning to build a deeper and wider network, and simultaneously using a multi-scale filter bank to jointly exploit spectral and spatial information of HSIs. The next is the CNN with pixel-pair features (CNN-PPF) \cite{Li2017CNNPPF}, which is a spectral-feature based method, using a series of pixel pairs as inputs. Similar to CNN-PPF, the siamese convolutional neural network (S-CNN) \cite{Liu2018SCNN} also took pairs of samples as inputs, yet used the pairs of patches to extract deep features, on which the support vector machine (SVM) was utilized for classification purposes. The last one is a 3-D CNN proposed in \cite{Plaza20183DCNN}, which actually extracted the spectral-spatial information with 2-D kernels. As indicated in \cite{Plaza20183DCNN}, keeping all the bands of an HSI as inputs could provide CNNs with full potential in spectral information mining, following which we also fed all the bands into the networks in our experiments.

Furthermore, two traditional supervised classification algorithms were implemented on hand-crafted Gabor features. The first one is the multinomial logistic regression (MLR) via variable splitting and augmented Lagrangian algorithm \cite{Bioucas2009LORSAL}, and the other one is the probabilistic SVM, which estimates the probabilities via combining pairwise comparisons \cite{Lin2007ProbSVM}. Both methods have been proven successful when dealing with high-dimensional data.

\subsection{Classification Results} \label{sec:exp:classif}

In the following, we describe the obtained experimental results in detail.

\subsubsection{Experiments with the Pavia University Scene}

\begin{figure*}[thp]
\scriptsize
\centering
  \includegraphics[height=1.55in]{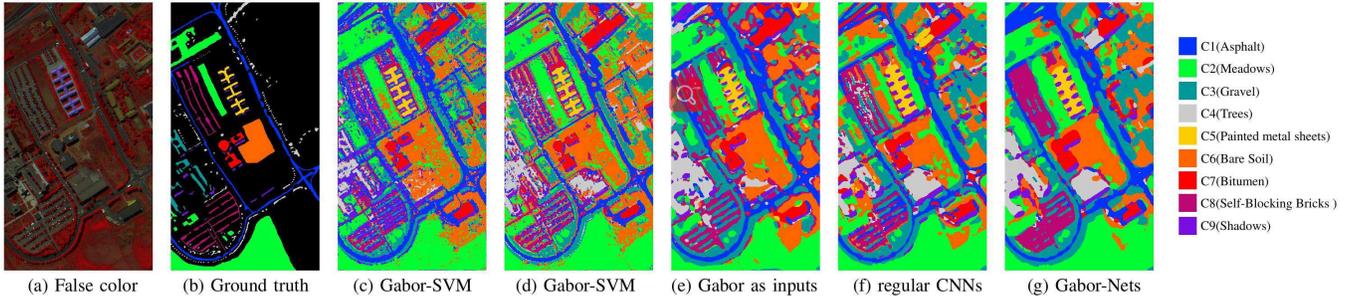}\\
\caption{The false color map, the ground truth, and the classification maps obtained from 5 Monte Carlo runs using 50 training samples per class for the Pavia University scene.}
\label{fig:pu:maps}
\end{figure*}

This scene is a benchmark hyperspectral dataset for classification, which was collected by the Reflective Optics Imaging Spectrometer System (ROSIS) sensor over the urban area of the University of Pavia, Italy, in 2003 (see Fig. \ref{fig:pu:maps} (a)). The image contains $610\times340$ samples with a spatial resolution of 1.3m, and 103 spectral bands ranging from 0.43 to 0.86 $\mu$m. The ground truth contains 42776 labeled samples within 9 classes of interest, where the numbers of samples corresponding to C1-C9 are 6631, 18649, 2099, 3064, 1345, 5029, 1330, 3682 and 947, respectively. The training samples were randomly selected from each class, and the rest were taken for test. To test the performance of Gabor-Nets with a small number of training samples, we evaluated their performance using 50, 100, and 200 training samples per class, respectively. We argue that 50 training samples per class is not a small training set for usual methods, but for deep learning based methods this number of training samples is actually limited with respect to the large number of model parameters. The patch size and and the kernel size were empirically set to 15 and 5, respectively. The regular CNNs and Gabor-Nets were constructed using two CV Blocks and one FC Block. In order to guarantee statistical significance, the results are reported by averaging five Monte Carlo runs corresponding to independent training sets.

First of all, we report the test accuracies obtained with different numbers of training samples for the Pavia University scene in Table \ref{table:pu:acc}. As can be observed, the proposed Gabor-Nets obtained very competitive results when compared to the other tested methods. The improvements were quite significant, especially in the case of 50 training samples per class. The 3-D CNN, Gabor as inputs and regular CNNs could obtain very close results to Gabor-Nets under the circumstance of 200 training samples per class. Nevertheless, Gabor-Nets outperformed 3-D CNN, Gabor as inputs and regular CNNs with accuracy gains of 5\%, 2\% and 10\%, respectively, when using 50 training samples per class. Furthermore, we implemented a data augmentation strategy for regular CNNs and Gabor-Nets, by mirroring each of them across the horizontal, vertical, and diagonal axes, respectively \cite{Lee2017DCCNN}. As shown in Table \ref{table:pu:acc}, the data augmentation strategy benefitted regular CNNs much more than Gabor-Nets, indicating that Gabor-Nets were less negatively affected by a lack of training samples than regular CNNs. This is expected since Gabor-Nets involve much less free parameters than regular CNNs. Besides, Gabor filters are able to achieve optimal resolution in both space and frequency domains, thus suitable for feature extraction purposes. Therefore, Gabor-Nets based on Gabor kernels can still yield representative features to some extent with limited training samples.

\begin{table}[htp]
\scriptsize
\begin{center}
\begin{tabular}{l|p{1.4cm}<{\centering}|p{1.4cm}<{\centering}|p{1.4cm}<{\centering}}
\hline
\hline
$\sharp$ Train                                          & 50/class                   & 100/class                  & 200/class   \\
\hline\hline
Gabor-MLR                               				& 92.41$\pm$0.66             & 94.96$\pm$0.30             & 96.73$\pm$0.25 \\
Gabor-SVM                               				& 90.99$\pm$0.72             & 92.93$\pm$0.91             & 95.79$\pm$0.38 \\
\hline
CDCNN \cite{Lee2017DCCNN}$^\ast$         				& -                          & -                          & 95.97 \\
CNN-PPF\cite{Li2017CNNPPF}$^\ast$        				& -                          & -                          & 96.48 \\
S-CNN\cite{Liu2018SCNN}$^\ast$        		     		& -                          & -                          & 99.08 \\
3-D CNN \cite{Plaza20183DCNN}$^\ast$       				& 90.22$\pm$1.78             & 94.37$\pm$1.10             & 98.06$\pm$0.13 \\
Gabor as inputs \cite{Chen2017InputGaborHSI}            & 93.20$\pm$1.55             & 96.14$\pm$0.75             & 98.79$\pm$0.29 \\
\hline
Regular CNNs                        		            & 85.52$\pm$1.51             & 95.43$\pm$0.60             & 98.12$\pm$0.18 \\
\textbf{Gabor-Nets}                               	    & \textbf{95.91$\pm$1.53}    & \textbf{98.40$\pm$0.31}    & \textbf{99.22$\pm$0.19} \\
CNNs+Aug.                                               & 94.67$\pm$1.07             & 97.53$\pm$0.20             & 99.10$\pm$0.18 \\
\textbf{Gabor-Nets+Aug.}                                & \textbf{97.28$\pm$1.09}    & \textbf{98.65$\pm$0.38}    & \textbf{99.48$\pm$0.06} \\
\hline\hline
\end{tabular}
\end{center}
\caption{Test accuracies (\%) obtained from 5 Monte Carlo runs with 50, 100 and 200 training samples per class, respectively, for the ROSIS Pavia University scene, where Aug. refers to an implementation with the data augmentation strategy, and $^\ast$ marks the implementations carried out by other authors.}
\label{table:pu:acc}
\end{table}

\begin{figure}[htp]
\scriptsize
\centering
  \begin{tabular}{p{3.8cm}<{\centering}p{3.8cm}<{\centering}}
   \includegraphics[height=1.2in]{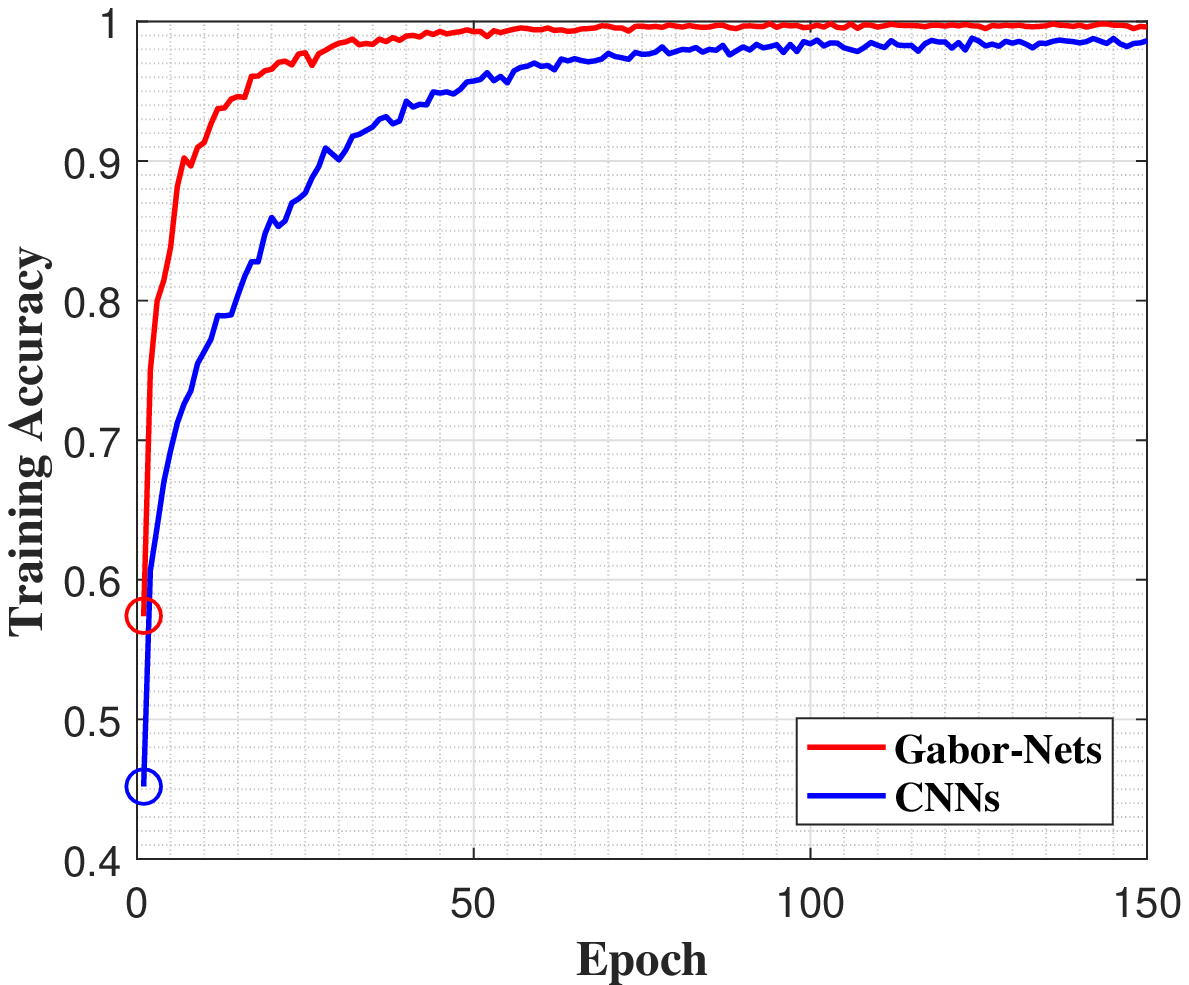}&
   \includegraphics[height=1.2in]{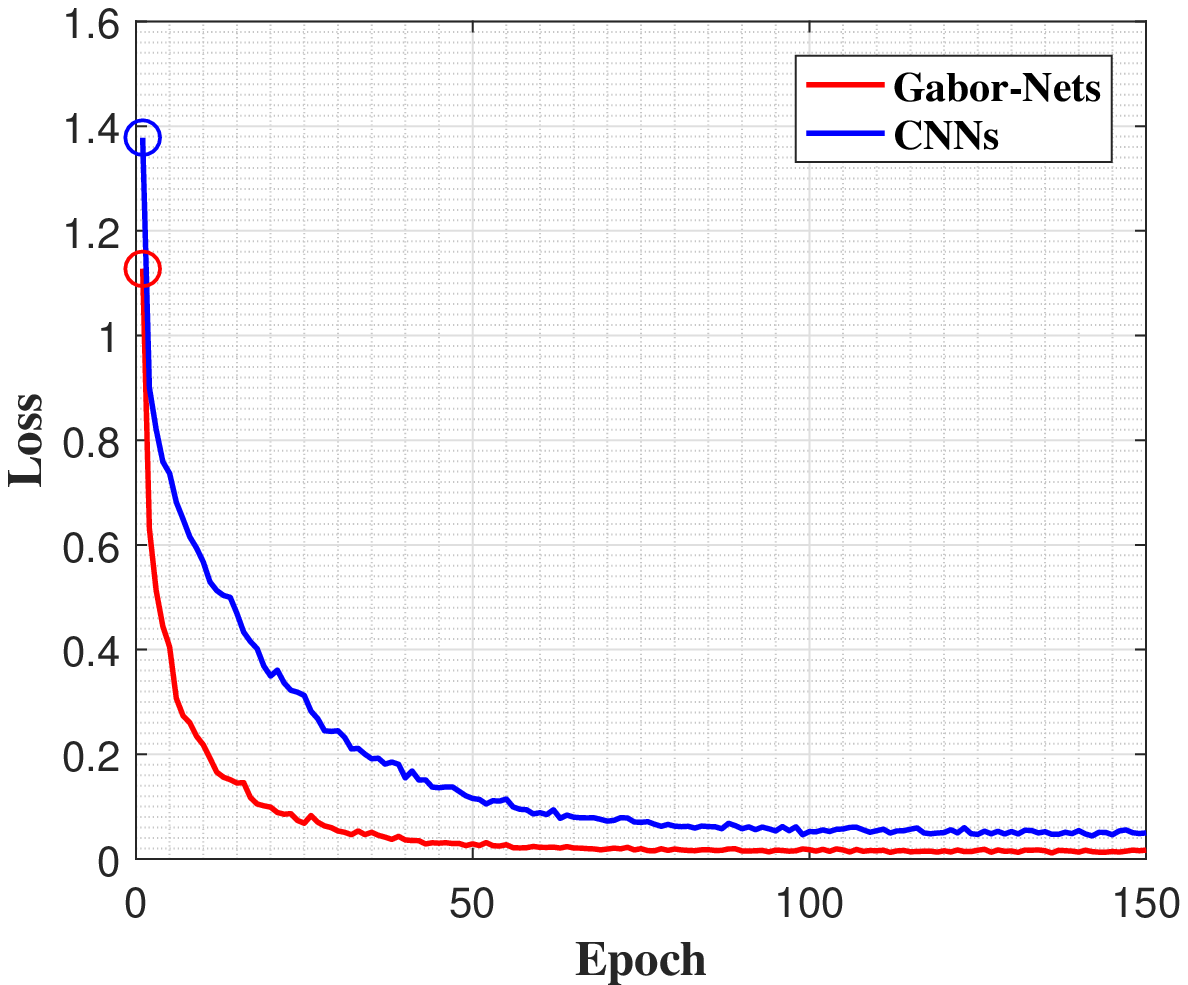}\\
  (a) Training accuracy & (b) Loss    \\
  \end{tabular}
\caption{(a) Training accuracies and (b) losses as functions of the number of epochs obtained using 100 training samples per class by Gabor-Nets and the regular CNNs, respectively. The initial values are marked with circles.}
\label{fig:acc:process}
\end{figure}

Fig. \ref{fig:acc:process} plots the training accuracies and losses obtained by the Regular CNNs and Gabor-Nets in the first 150 epochs. It can be seen that Gabor-Nets initially yielded a higher training accuracy and a smaller loss, and then converged faster, which indicates that Gabor-Nets constructed by Gabor kernels are able to constrain the solution space of CNNs, thus playing a positive role in the learning of CNNs.

Some of the classification maps obtained using 50 training samples per class are shown in Fig. \ref{fig:pu:maps}. It can be seen that the classification map obtained by Gabor-Nets is smoother than those obtained by other methods. In contrast, the maps obtained by traditional methods are negatively affected by the appearance of noises. It is known that CNNs extract features via multi-layer convolutions, while traditional shallow filtering methods convolve the image using a single-layer strategy, which makes CNNs have a better ability to remove noises. However, this also tends to make CNNs over-smooth HSIs sometimes, which leads to the information loss especially of small ground objects.

\begin{figure}[thp]
\scriptsize
\centering
  \begin{tabular}{c}
  \includegraphics[width=1.8in]{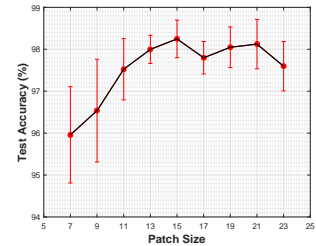}
  \end{tabular}
\caption{Test accuracies (along with standard deviations) as a function of patch sizes for the Pavia University scene using 100 training samples per class.}
\label{fig:pu:patch}
\end{figure}

Then, we investigate the relationship between the patch size and the classification performance of Gabor-Nets. We set the patch sizes varying from 7 to 23 with an increasing interval of 2 pixels, and illustrate the obtained test accuracies along with standard deviations in Fig. \ref{fig:pu:patch}. It can be observed that very small patches had a negative effect on the classification accuracies and the robustness of Gabor-Nets. As the patch size increased, Gabor-Nets performed better. However, when the patch size became very large, the performance decreased again. We argue that the patches can be regarded as a series of local spatial dependency systems \cite{He2018Review}, using the patch size to define the neighborhood coverage. According to \textit{Tobler's first law of geography}, the similarity between two objects on the same geographical surface has an inverse relationship with their distance. Therefore, those samples located at a distance away from the central one are not helpful (and even will confuse the classifier), whereas small patches are unable to provide relevant information, thus limiting the potential of networks.

\begin{table*}[ht]
\scriptsize
\begin{center}
\begin{tabular}{p{2.0cm}<{\centering}|p{2.4cm}<{\centering}|p{2.4cm}<{\centering}|p{2.4cm}<{\centering}|p{2.6cm}<{\centering}|p{3.2cm}<{\centering}}
\hline\hline
\multicolumn{2}{c|}{$\sharp$CV Blocks} &  1 & 2 & 3 & 4 \\
\hline
\multicolumn{2}{c|}{$\sharp$Kernels per Conv layer} & 16-16 & 16-16-32-32 & 16-16-32-32-64-64 & 16-16-32-32-64-64-128-128 \\
\hline\hline
Regular CNNs & Test Accuracies ($\sharp$para) & 92.47$\pm$1.68 (48K) & 95.43$\pm$0.60 (89K) & 92.96$\pm$1.44 (249K) & 86.93$\pm$3.28 (890K)  \\
\hline
Gabor-Nets & Test Accuracies ($\sharp$para) & 96.37$\pm$1.10 (8K)  & 98.40$\pm$0.31 (17K) & 98.46$\pm$0.22 (48K) & 98.19$\pm$0.74 (172K)  \\
\hline\hline
\end{tabular}
\end{center}
\caption{Test accuracies (\%) along with standard deviations (\%) obtained using 100 training samples per class from 5 Monte Carlo runs by the regular CNNs and Gabor-Nets constructed with different numbers of CV Blocks for the Pavia University scene. The numbers of trainable parameters are listed in brackets.}
\label{table:pu:moreBlocks}
\end{table*}

Finally, we test the performance of Gabor-Nets with different numbers of CV Blocks. From Table \ref{table:pu:moreBlocks} we can observe that Gabor-Nets exhibit better robustness when using different numbers of CV Blocks. In this case, Gabor-Nets were able to achieve more reliable results with more CV Blocks added, whereas the performance of regular CNNs degraded severely, as a result of overfitting caused by a sharp rise in parameter numbers. Additionally, both Gabor-Nets and regular CNNs performed worse when the number of CV Blocks decreased to 1, partly due to the decline in the representation ability of networks. Yet the drop in test accuracies of Gabor-Nets is around 1\% less than that of regular CNNs, which suggests the superiority of Gabor-Nets employing the Gabor \textit{a priori}.

\begin{figure}[thp]
\scriptsize
\centering
    \includegraphics[height=1.85in]{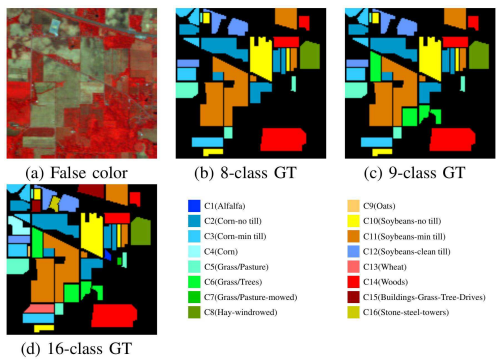}\\
\caption{The false color map along with the ground truth (GT) of 8, 9 and 16 classes, respectively, for the Indian Pines scene.}
\label{fig:indian:data}
\end{figure}

\subsubsection{Experiments with the Indian Pines Scene}

The second dataset used in our experiments is the well-known Indian Pines scene, collected over a mixed agricultural/forest area in North-western Indiana, USA, by the Airborne Visible Infrared Imaging Spectrometer (AVIRIS) sensor in 1992. This scene is composed of 220 spectral bands with wavelength varying from 0.4 to 2.5 $\mu$m, and 145$\times$145 pixels with a spatial coverage of 20m$\times$20m. In our experiments, we removed 20 bands due to noises and water absorption, resulting in 200 bands. This scene is challenging for traditional HSI classification methods due to the fact that most of the samples are highly mixed. As shown in Fig. \ref{fig:indian:data} (d) and Table \ref{table:indian:16tnum}, the available ground truth (GT) contains 10249 labeled samples belonging to 16 unbalanced classes. To tackle this problem, Liu \textit{et al.} \cite{Liu2018SCNN} and Li \textit{et al.} \cite{Li2017CNNPPF} removed C1, C4, C7, C9, C13, C15, and C16 from the original GT, leaving 9 classes. Lee \textit{et al.} \cite{Lee2017DCCNN} removed C6 besides the above seven classes, leaving 8 classes. Furthermore, Paoletti \textit{et al.} \cite{Plaza20183DCNN} balanced the number of training samples of each class in accordance with their sample sizes via a stratified sampling strategy. For comparison purposes, we considered all the three circumstances in our experiments, and randomly selected 50, 100, and 200 samples per class for training, leaving the remains for test. Specifically, Table \ref{table:indian:16tnum} presents the numbers of training and test samples using the 16-class GT \cite{Plaza20183DCNN}. We utilized a similar network architecture to the previous one, i.e., two CV Blocks and one FC Block, with the patch size of 15 and the filter size of 5. We conducted five Monte Carlo runs and reported the average results in the following.

\begin{table}[ht]
\scriptsize
\begin{center}
\begin{tabular}
{c|c|c|c|c|c|c|c}
\hline\hline
\multicolumn{1}{c|}{\multirow{2}{*}{Class}} &  \multicolumn{1}{c|}{\multirow{2}{*}{$\sharp$Sample}} & \multicolumn{2}{c|}{\multirow{1}{*}{50/class}} & \multicolumn{2}{c|}{\multirow{1}{*}{100/class}} & \multicolumn{2}{c}{\multirow{1}{*}{200/class}} \\
\cline{3-8}
& & $\sharp$Train & $\sharp$Test & $\sharp$Train & $\sharp$Test & $\sharp$Train & $\sharp$Test\\
\hline\hline
C1    & 46       & 33 &    13 & 33  &    13 & 33  &    13 \\
C2    & 1428     & 50 &  1378 & 100 &  1328 & 200 &  1228 \\
C3    & 830      & 50 &   780 & 100 &   730 & 200 &   630 \\
C4    & 237      & 50 &   187 & 100 &   137 & 181 &    56 \\
C5    & 483      & 50 &   433 & 100 &   383 & 200 &   283 \\
C6    & 730      & 50 &   680 & 100 &   630 & 200 &   530 \\
C7    & 28       & 20 &     8 & 20  &     8 & 20  &     8 \\
C8    & 478      & 50 &   428 & 100 &   378 & 200 &   278 \\
C9    & 20       & 14 &     6 & 14  &     6 & 14  &     6 \\
C10   & 972      & 50 &   922 & 100 &   872 & 200 &   772 \\
C11   & 2455     & 50 &  2405 & 100 &  2355 & 200 &  2255 \\
C12   & 593      & 50 &   543 & 100 &   493 & 200 &   393 \\
C13   & 205      & 50 &   155 & 100 &   105 & 143 &    62 \\
C14   & 1265     & 50 &  1215 & 100 &  1165 & 200 &  1065 \\
C15   & 386      & 50 &   336 & 100 &   286 & 200 &   186 \\
C16   & 93       & 50 &    43 & 75  &    18 & 75  &    18 \\
\hline
Total                             & 10249    &717 &  9532 &1342 &  8907 &2466 &  7783 \\
\hline\hline
\end{tabular}
\end{center}
\caption{The numbers of training and test samples per class using the ground truth of 16 classes for the Indian Pines scene.}
\label{table:indian:16tnum}
\end{table}

Tables \ref{table:ip8c:acc}-\ref{table:ip16c:acc} list the test accuracies obtained using the ground truth of 8, 9 and 16 classes, respectively. Clearly, Gabor-Nets outperformed the other methods in all the considered cases, especially when only 50 training samples per class were utilized, indicating that Gabor-Nets have a capability to deal with limited training samples. Additionally, Gabor-Nets yielded better classification results than Gabor as inputs, from which we can infer that Gabor-Nets, via adjusting the Gabor parameters in a data-driven manner, were able to generate more effective features than hand-crafted Gabor filters. Furthermore, most deep learning based methods outperformed the traditional ones, showing the potential of CNNs in HSI classification tasks.

\begin{figure*}[thp]
\scriptsize
\centering
    \includegraphics[height=1.00in]{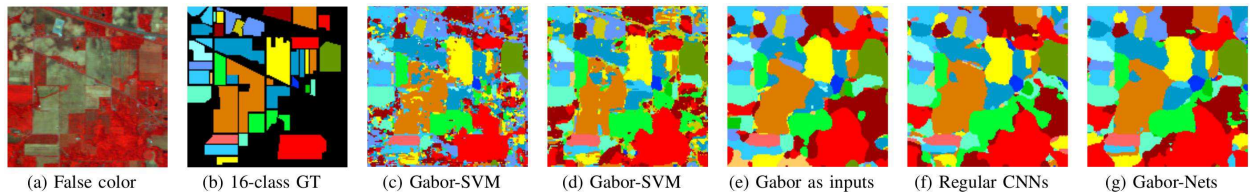}\\
\caption{Classification maps obtained using the ground truth (GT) of 16 classes and 50 randomly selected training samples per class (the actual numbers of samples from each class are shown in Table \ref{table:indian:16tnum}) for the Indian Pines scene.}
\label{fig:indian:maps}
\end{figure*}

\begin{table}[thp]
\scriptsize
\begin{center}
\begin{tabular}{l|p{1.4cm}<{\centering}|p{1.4cm}<{\centering}|p{1.4cm}<{\centering}}
\hline\hline
$\sharp$Train                                & 50/class          & 100/class         & 200/class\\
\hline\hline
Gabor-MLR                               	 & 86.70$\pm$0.84    & 93.63$\pm$0.57    & 97.04$\pm$0.39\\
Gabor-SVM                               	 & 83.79$\pm$0.98    & 91.08$\pm$0.54    & 96.22$\pm$0.68\\
\hline
CDCNN \cite{Lee2017DCCNN}$^\ast$         	 & -                 & -                 & 93.61$\pm$0.56\\	
Gabor as inputs\cite{Chen2017InputGaborHSI}  & 93.45$\pm$1.11    & 97.34$\pm$1.04    & 98.86$\pm$0.31\\
\hline
Regular CNNs                            	 & 91.83$\pm$3.31    & 96.50$\pm$0.78    & 99.12$\pm$0.31\\
\textbf{Gabor-Nets}                          & \textbf{94.33$\pm$0.42}    & \textbf{97.58$\pm$0.43}    & \textbf{99.24$\pm$0.33}\\
\hline\hline
\end{tabular}
\end{center}
\caption{Test accuracies (\%) obtained from 5 Monte Carlo runs with 50, 100, and 200 training samples per class, respectively, randomly selected from the ground truth (GT) of 8 classes for AVIRIS Indian Pines scene.}
\label{table:ip8c:acc}
\end{table}

\begin{table}[thp]
\scriptsize
\begin{center}
\begin{tabular}{l|p{1.4cm}<{\centering}|p{1.4cm}<{\centering}|p{1.4cm}<{\centering}}
\hline\hline
$\sharp$Train                                & 50/class                   & 100/class                     & 200/class \\
\hline\hline
Gabor-MLR                               	 & 88.34$\pm$1.08             & 93.78$\pm$0.61                & 96.99$\pm$0.41\\
Gabor-SVM                               	 & 84.39$\pm$0.51             & 91.44$\pm$0.58                & 96.11$\pm$0.51\\
\hline	
CNN-PPF\cite{Li2017CNNPPF}$^\ast$        	 & -                          & -    & 94.34\\		
S-CNN\cite{Liu2018SCNN}$^\ast$        		 & -                          & -    & 99.04\\	
Gabor as inputs\cite{Chen2017InputGaborHSI}  & 93.65$\pm$1.07             & 97.55$\pm$0.31                & 98.84$\pm$0.31\\
\hline
Regular CNNs                            	 & 92.41$\pm$2.45             & 96.42$\pm$0.83                & 98.73$\pm$0.44\\
\textbf{Gabor-Nets}                          & \textbf{94.76$\pm$0.46}    & \textbf{97.54$\pm$0.16}       & \textbf{99.05$\pm$0.19}\\
\hline\hline
\end{tabular}
\end{center}
\caption{Test accuracies (\%) obtained from 5 Monte Carlo runs with 50, 100, and 200 training samples per class, respectively, randomly selected from the ground truth (GT) of 9 classes for the Indian Pines scene.}
\label{table:ip9c:acc}
\end{table}

\begin{table}[thp]
\scriptsize
\begin{center}
\begin{tabular}{l|p{1.4cm}<{\centering}|p{1.4cm}<{\centering}|p{1.4cm}<{\centering}}
\hline\hline
$\sharp$Train                                & 50/class          & 100/class         & 200/class \\
\hline\hline
Gabor-MLR                               	 & 87.63$\pm$0.79    & 94.05$\pm$0.71    & 97.18$\pm$0.32 			\\
Gabor-SVM                               	 & 85.15$\pm$0.43    & 92.23$\pm$0.36    & 96.18$\pm$0.37 			\\
\hline	
3-D CNN\cite{Plaza20183DCNN}$^\ast$       	 & 88.78$\pm$0.78    & 95.05$\pm$0.28    & 98.37$\pm$0.17 			\\	
Gabor as inputs\cite{Chen2017InputGaborHSI}  & 93.29$\pm$1.09    & 96.91$\pm$0.63    & 98.67$\pm$0.39 			\\
\hline
Regular CNNs                            	 & 92.74$\pm$0.67    & 96.42$\pm$0.47    & 98.28$\pm$0.58 			\\
\textbf{Gabor-Nets}                          & \textbf{94.05$\pm$0.79}    & \textbf{97.01$\pm$0.52}    & \textbf{98.75$\pm$0.38} 			\\
\hline\hline
\end{tabular}
\end{center}
\caption{Test accuracies (\%) obtained from 5 Monte Carlo runs with 50, 100, and 200 training samples per class, respectively, randomly selected from the ground truth (GT) of 16 classes for the Indian Pines scene.}
\label{table:ip16c:acc}
\end{table}

Fig. \ref{fig:indian:maps} shows some of the classification maps obtained with 50 training samples per class using the 16-class GT. As illustrated, the assignments by Gabor-Nets are more accurate, and the corresponding classification map looks smoother. However, the maps by CNN methods are somewhat over-smoothed on this scene, partly due to their multi-layer feature extraction strategy, which makes CNNs prone to over-smooth hyperspectral images, especially when the interclass spectral variability is low, as in the case of the Indian Pines scene.

\subsubsection{Experiments with the Houston Scene}

\begin{figure}[thp]
\scriptsize
\centering
    \includegraphics[width=7.8cm]{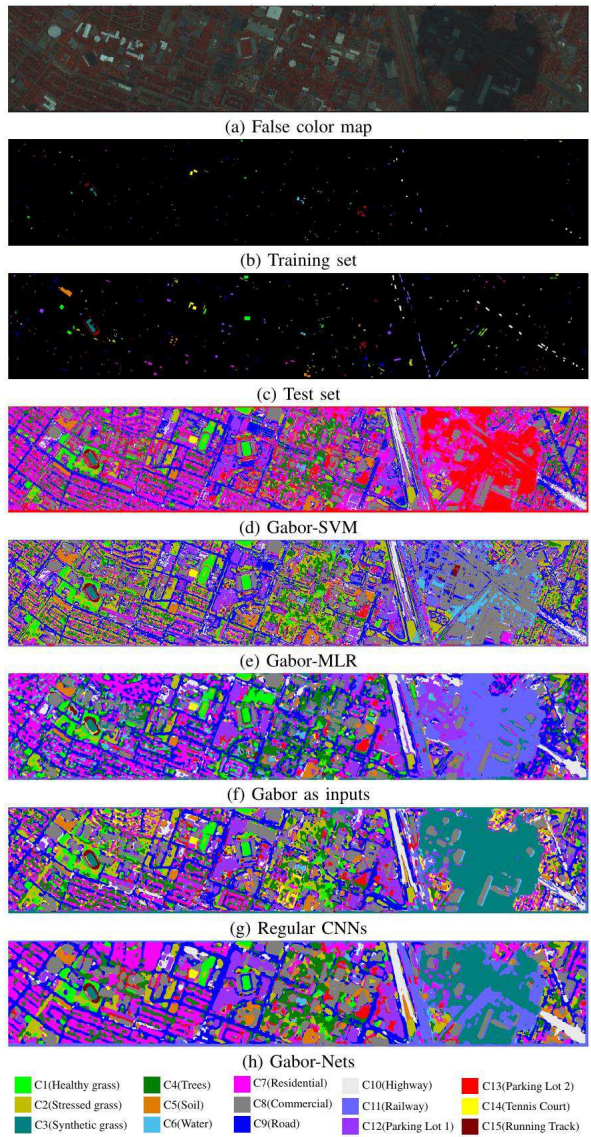}\\
\caption{The false color map, training set, test set, and classification maps for the Houston scene.}
\label{fig:houston:maps}
\end{figure}

The Houston scene was acquired by the Compact Airborne Spectrographic Imagery from the ITRES company (ITRES-CASI 1500) over the area of University of Houston campus and the neighboring urban area in 2012. It was first known and distributed as the hyperspectral image provided for the 2013 IEEE Geoscience and Remote Sensing Society (GRSS) data fusion contest. This scene is composed of 349$\times$1905 pixels at a spatial resolution of 2.5m and 144 spectral bands ranging from 380nm to 1050nm. The public ground truth contains 15029 labeled samples of 15 classes, including 2832 training samples (198, 190, 192, 188, 186, 182, 196, 191, 193, 191, 181, 192, 184, 181 and 187 corresponding to C1-C15, respectively) and 12197 test samples (1053, 1064, 505, 1056, 1056, 143, 1072, 1053, 1059, 1036, 1054, 1041, 285, 247 and 473 corresponding to C1-C15, respectively) as shown in Fig. \ref{fig:houston:maps}. 
This dataset is a typical urban scene with a complex spatial appearance containing many natural or artificial ground fabrics, based on which, we utilized three CV Blocks and one FC Block to mine deeper feature representations, and empirically set the patch size and filter size as 13 and 3, respectively. We utilized the publicly available training set and repeated the tested CNN-based methods five times w.r.t different random initializations.

First and foremost, we quantitatively evaluate the classification performance in Table \ref{table:houston:acc}, where the proposed Gabor-Nets obtained the highest test accuracies. However, other deep learning based methods could not outperform the traditional ones as much as they did on the previous scenes. These observations indicate that Gabor-Nets also exhibit potential when dealing with complex urban scenarios.

\begin{table}[htp]
\scriptsize
\begin{center}
\begin{tabular}{p{3.5cm}|p{1.8cm}<{\centering}}
\hline
\hline
\hspace{0.5cm}Methods                                                                                 & Test accuracies   \\
\hline\hline
\hspace{0.5cm}Gabor-MLR                               				                                  & 79.86    \\
\hspace{0.5cm}Gabor-SVM                               				                                  & 79.38    \\
\hline
\hspace{0.5cm}CNN-PPF\cite{Li2017CNNPPF}\cite{Li201DL-HSI-ClassificationReview}$^\ast$        	      & 81.38                 \\
\hspace{0.5cm}S-CNN\cite{Liu2018SCNN}\cite{Li201DL-HSI-ClassificationReview}$^\ast$        		      & 82.34                 \\
\hspace{0.5cm}Gabor as inputs \cite{Chen2017InputGaborHSI}                                            & 79.43$\pm$0.91    \\
\hline
\hspace{0.5cm}Regular CNNs                        		                                              & 78.55$\pm$0.99    \\
\hspace{0.5cm}\textbf{Gabor-Nets}                               	                                  & \textbf{85.57$\pm$1.18}   \\
\hline\hline
\end{tabular}
\end{center}
\caption{Test accuracies (\%) obtained from 5 Monte Carlo runs using the public training and test sets for the Houston scene, where $^\ast$ marks the implementations carried out by other authors.}
\label{table:houston:acc}
\end{table}

A visual comparison can be found in Fig. \ref{fig:houston:maps}, where the classification map generated by Gabor-Nets looks smoother than the others, also with clear roads. Furthermore, more details below the cloud are revealed on the map by Gabor-Nets than those by Gabor as inputs and regular CNNs, which suggests the effectiveness of the proposed Gabor kernels. Besides, the maps by traditional methods are severely affected by the appearance of noises, though they yielded very close statistic results to CNNs and Gabor as inputs.

\begin{figure}[thp]
\scriptsize
\centering
  \begin{tabular}{c}
  \includegraphics[width=1.8in]{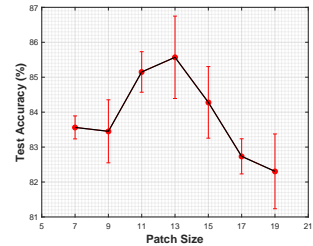}
  \end{tabular}
\caption{Test accuracies (along with standard deviations) as a function of patch sizes for the Houston scene using the public training samples.}
\label{fig:houston:patch}
\end{figure}

Next, we evaluated the performance of Gabor-Nets with varying patch sizes in Fig. \ref{fig:houston:patch}. Similar to the experiments on the Pavia University scene, too small and too big patches both had a negative effect on the performance of Gabor-Nets. This scene is more sensitive to big patches. As mentioned before, this scene is quite complex spatially. As a result, big patches tend to damage its underlying spatial structure.

Finally, we investigate the relationship between the number of CV Blocks and the classification performance of the proposed Gabor-Nets. As illustrated in Table \ref{table:houston:moreBlocks}, for all the considered architectures, the number of required parameters of Gabor-Nets was only around half of that of regular CNNs. Also, Gabor-Nets performed better than regular CNNs, no matter what architecture was utilized. The Gabor-Nets were able to maintain its superiority as more CV Blocks were involved into the architectures, whereas an obvious degradation can be observed in the obtained test accuracies of regular CNNs when the number of CV Blocks increased from 3 to 4.

\begin{table*}[ht]
\scriptsize
\begin{center}
\begin{tabular}{p{2.0cm}<{\centering}|p{2.4cm}<{\centering}|p{2.4cm}<{\centering}|p{2.4cm}<{\centering}|p{2.6cm}<{\centering}|p{3.2cm}<{\centering}}
\hline\hline
\multicolumn{2}{c|}{$\sharp$CV Blocks} &  1 & 2 & 3 & 4 \\
\hline
\multicolumn{2}{c|}{$\sharp$Kernels per Conv layer} & 16-16 & 16-16-32-32 & 16-16-32-32-64-64 & 16-16-32-32-64-64-128-128 \\
\hline\hline
Regular CNNs & Test Accuracies ($\sharp$para) & 71.90$\pm$2.84 (24K) & 80.59$\pm$3.67 (40K) & 78.55$\pm$0.99 (103K) & 72.38$\pm$2.67 (351K)  \\
\hline
Gabor-Nets & Test Accuracies ($\sharp$para) & 77.37$\pm$1.65 (11K)   & 83.01$\pm$0.88 (20K) & 85.57$\pm$1.18 (51K) & 85.43$\pm$1.34 (177K)  \\
\hline\hline
\end{tabular}
\end{center}
\caption{Test accuracies (\%) along with standard deviations (\%) obtained from 5 Monte Carlo runs using the public training set by the regular CNNs and Gabor-Nets constructed using different numbers of CV Blocks for the Houston scene. The numbers of trainable parameters are listed in brackets.}
\label{table:houston:moreBlocks}
\end{table*}

\begin{figure*}[thp]
\scriptsize
\centering
  \includegraphics[height=1.75in]{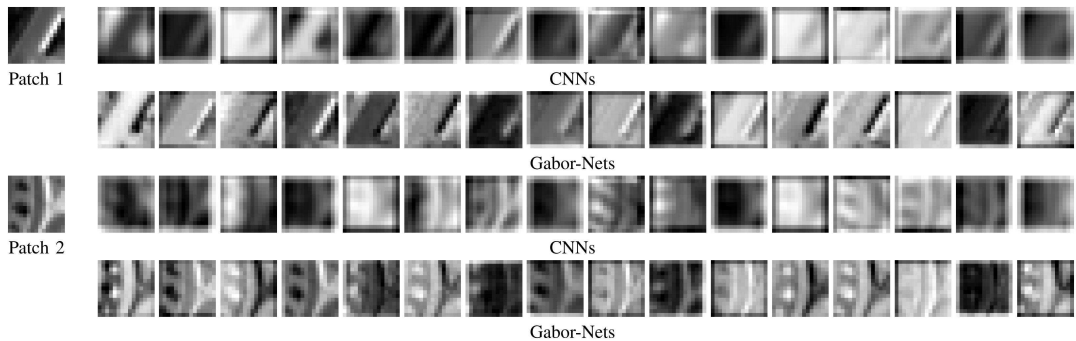}\\
\caption{Two patches along with their corresponding output features from the first layers of CNNs and Gabor-Nets, respectively, using 100 training samples per class for the Pavia University scene.}
\label{fig:pavia:FeatVisual}
\end{figure*}

\begin{table*}[htp]
  \scriptsize
  \centering
  \begin{tabular}{c|c|c|c|c|c}
  \hline\hline
  Random initializations & -- & a & b & c & abc\\
  \hline
  Test accuracies (\%)   & 96.37$\pm$1.10 & 96.07$\pm$1.37 & 95.27$\pm$1.05 & 88.07$\pm$3.63 & 90.17$\pm$0.84\\
  \hline\hline
  \end{tabular}
  \caption{Test accuracies obtained with different initializations, where "--" denotes the initialization scheme proposed in this work, otherwise the letters represent the cases of using certain random initialization(s) to replace the corresponding original initialization(s) in the proposed initialization scheme.}
  \label{table:OA:randInit}
\end{table*}

\subsection{Model Insight}

To further analyze the mechanism behind Gabor-Nets, we investigate some of the properties of Gabor-Nets. For illustrative purposes, we focus on the experiments with the Pavia University scene (using 100 randomly selected training samples per class without the augmentation strategy).

\subsubsection{Visualizations of first-layer features}

To help readers understand more about what Gabor-Nets learn at the bottom layers, we visualize their first-layer features along with those of regular CNNs in Fig. \ref{fig:pavia:FeatVisual}, which are extracted from two patches (15$\times$15) of the Pavia University scene. As illustrated, both regular CNNs and Gabor-Nets can extract features at certain orientations, which indicates that their low-layer features share some similar characteristics. However, the features extracted by regular CNNs are somewhat blurred and in various shapes. In contrast, the boundaries are depicted clearly on the feature maps by Gabor kernels, and each feature map reflects the information specific to an orientation and a frequency. This confirms that, although the variety of the features obtained by Gabor kernels is not as high as that obtained by regular kernels, Gabor-Nets can extract more compact and representative features, since the underlying features of HSIs are relatively simple, mainly composed of a series of geometrical and morphological features, in which case Gabor filters have been proven effective.

\subsubsection{Initialization scheme}

In this work, we design an initialization scheme in accordance to Gabor \textit{a priori} knowledge for Gabor-Nets to guarantee their performances. To verify the reliability of our initialization scheme, we conducted some experiments using the network architecture of one CV Block and one FC Block with random initializations of $\theta$s, $\omega$s and $\sigma$s for the Pavia University scene. Let $\tilde{\mu}$ and $\tilde{\sigma}$ denote the mean and the standard deviation of the normal distribution. The random initializations of $\theta$s, $\omega$s and $\sigma$s adopted in our experiments are as follows:

a. $\theta_0$s obeying a uniform distribution within $[0,2\pi)$;

b. $\omega_0$s obeying a normal distribution with $\tilde{\mu}$=$0$, $\tilde{\sigma}$=$\pi/4$;

c. $\sigma_0$s obeying a normal distribution with $\tilde{\mu}$=$0$, $\tilde{\sigma}$=$5/8$.

We utilize the above random initializations to replace the corresponding original initializations in the proposed initialization scheme, and report the obtained results in Table \ref{table:OA:randInit}, from which we can observe that the initialization scheme can guarantee Gabor-Nets to yield reliable results.

\begin{figure}[htp]
\scriptsize
\centering
  \begin{tabular}{p{3.8cm}<{\centering}p{3.8cm}<{\centering}}
  \includegraphics[height=1.2in]{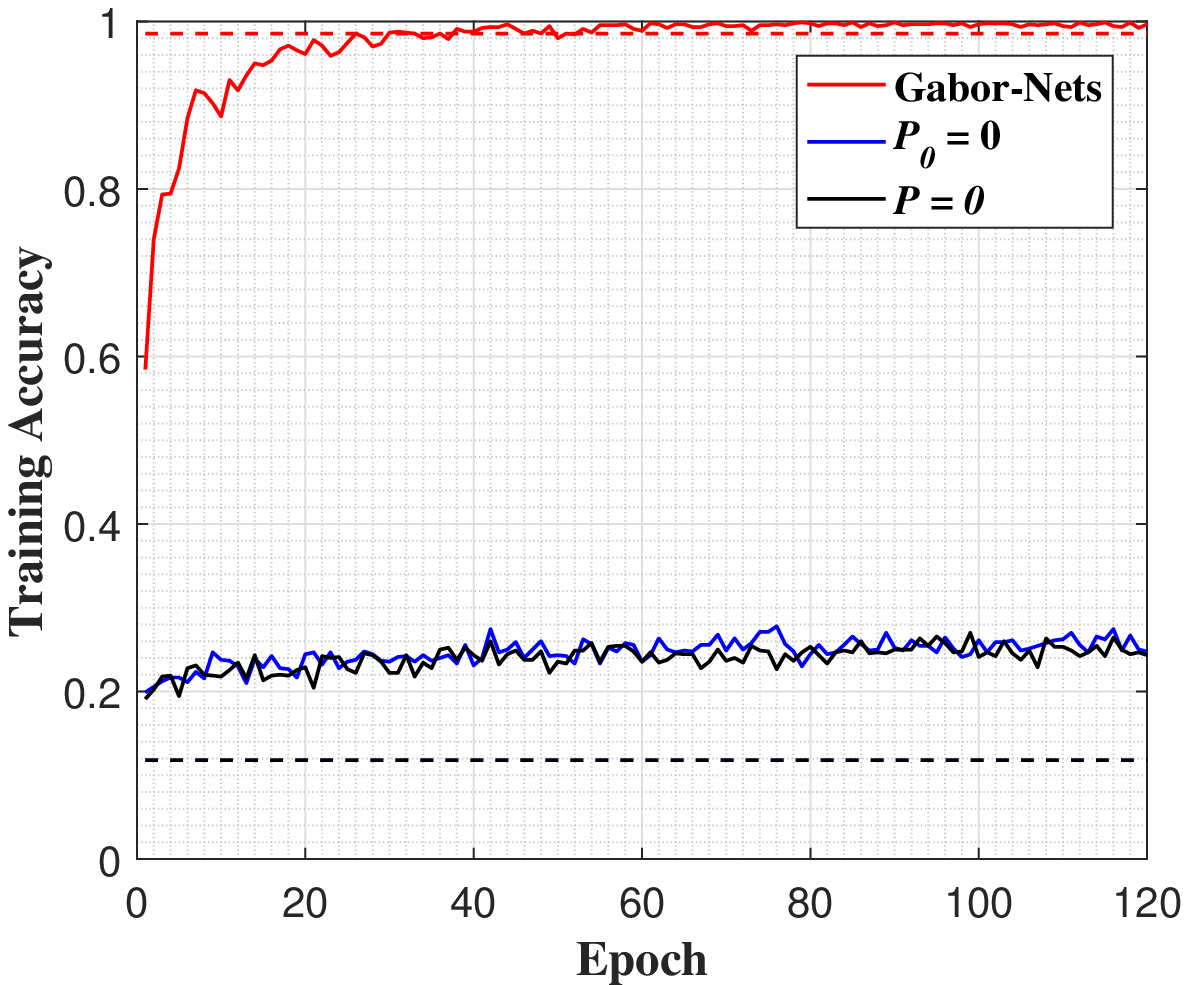}&
  \includegraphics[height=1.2in]{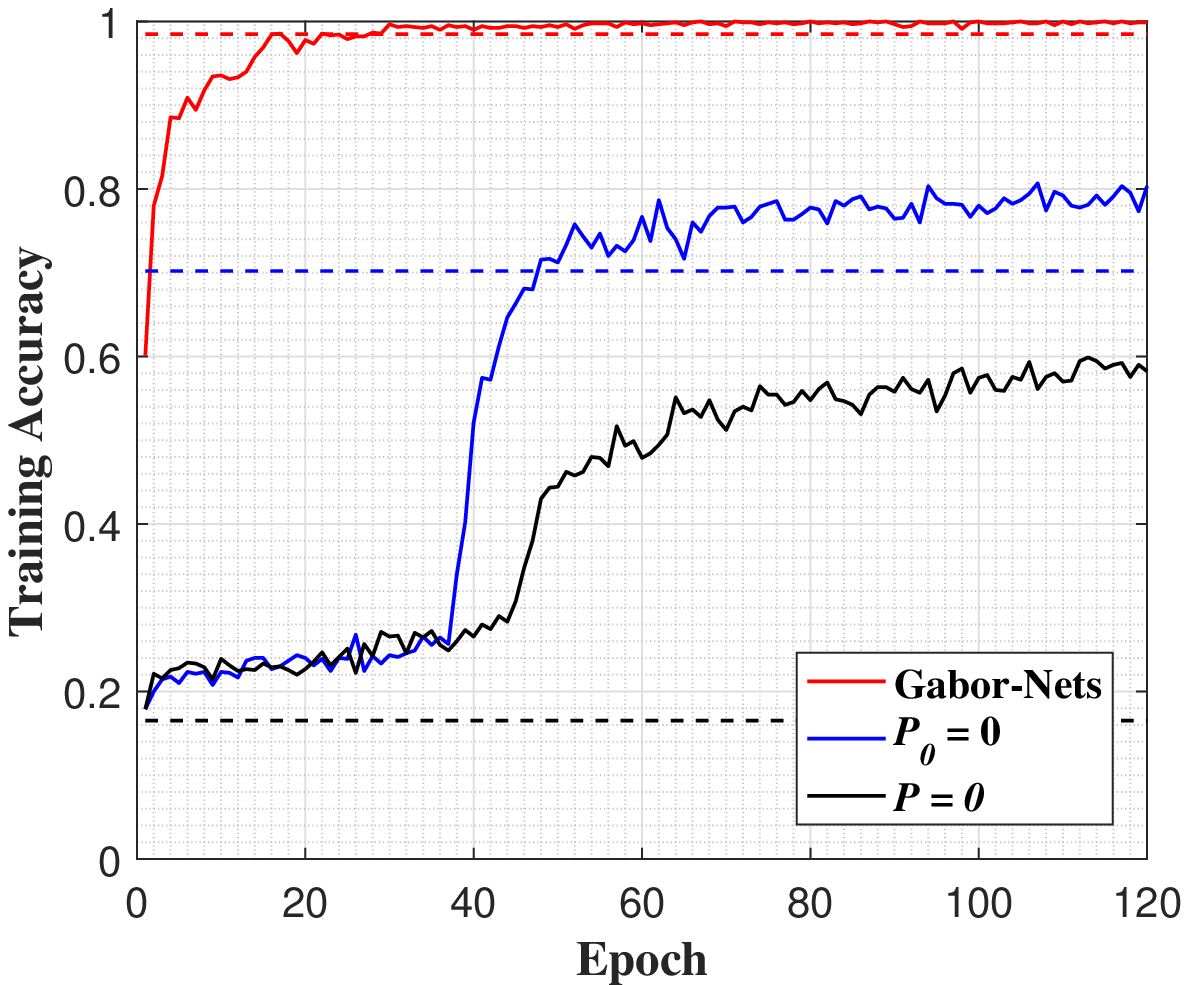}\\
  (a)  & (b)
  \end{tabular}
\caption{Training accuracies obtained with an initial learning rate of (a) 0.0076 and (b) 0.02, respectively, as functions of the number of epochs, where three types of Gabor-Nets are considered: the one that we proposed (red), the one with $P$s initialized to 0, i.e., $P_0=0$ (blue), and the one without $P$, i.e., $P=0$ (black). The dashed lines indicate their test accuracies.}
\label{fig:phase:process}
\end{figure}

\subsubsection{Phase Offsets}

As stated above, the kernel phase $P$ is crucial in Gabor-Nets, which controls the frequency characteristics of Gabor kernels. To test the role of $P$, we implemented two variants of Gabor-Nets, i.e., the one with all $P$s initialized to 0 ($P_0=0$); and the one without $P$ ($P=0$). Fig. \ref{fig:phase:process} shows the training accuracies obtained with initial learning rates of 0.0076 and 0.02, respectively, as functions of the number of epochs. Remarkably, randomly initializing $P$ in $[0,2\pi)$ can make Gabor-Nets achieve better performance and higher robustness with different learning rates in comparison to the two variants, where Gabor-Nets yielded results of around 98\% using both the considered initial learning rates. Quite opposite, the two variants performed much worse when using the initial learning rate of 0.0076. Recall that the gradient descend back propagation is a local search algorithm, in which reducing the learning rate will bring to smaller adjustments to the parameters, thus easily leading to the gradient vanishing phenomenon. Therefore, it can be inferred that Gabor-Nets with randomly initialized $P$s can resist against this phenomenon to some extent. The differences of test accuracies between the Gabor-Nets and two variants still exist when the learning rate increases, although the two variants performed better. Furthermore, the variant without $P$ yielded the worst results among the three models, which indicates that without the kernel phase term $P$, the ability of Gabor-Nets will be restricted since the frequency properties of Gabor kernels cannot adaptively follow the data. Regarding $P_0=0$, the fixed initialization will also harm the potential of Gabor-Nets due to a lack of diversity.

\begin{figure*}[htp]
\scriptsize
\centering
  \includegraphics[height=1.05in]{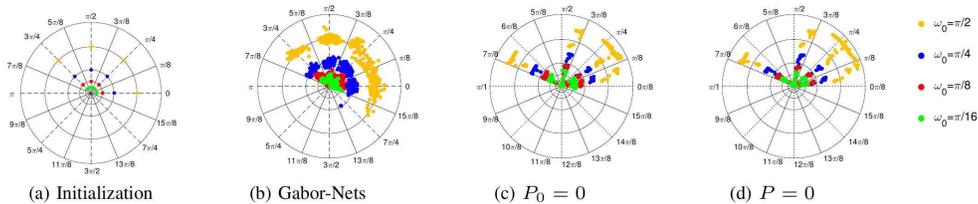}\\
\caption{Visualization of the initial and the finally learned frequencies of the Gabor kernels in the first layer. Each point $(\omega,\theta)$ corresponds to a learned kernel, constructed with the angle $\theta$ and the frequency magnitude $\omega$. The colored points represent the kernels with different $\omega_0$s. The circles (from inside out) represent the cases where $\omega$ is $\pi/16$, $\pi/8$, $\pi/4$,  $\pi/2$, and $2.4$, respectively. Notice that all the learned frequencies in the cases (b)-(d) used the same initialization shown in (a). (Best viewed in color.)}
\label{fig:phase:freq}
\end{figure*}

In another experiment, we investigate the learned angular frequencies of Gabor-Nets and the two variants. Fig. \ref{fig:phase:freq} shows the finally learned frequencies of Gabor kernels in the first layer (in terms of their angles and magnitudes), with an initial learning rate of 0.0076. Noticeably, the learned frequencies of Gabor-Nets tend to cover the whole semicircle region, while those of the two variants are only distributed at some local narrow regions, i.e., their $\theta$s and $\omega$s changed rarely in the learning process, which suggests that the two variants suffered from the gradient vanishing problem. Thus, from these results, we can infer that Gabor-Nets with randomly initialized kernel phases could resist against the gradient vanishing problem to some extent and positively affect the learning process of other parameters.

\begin{figure*}[htp]
\scriptsize
\centering
  \includegraphics[height=2in]{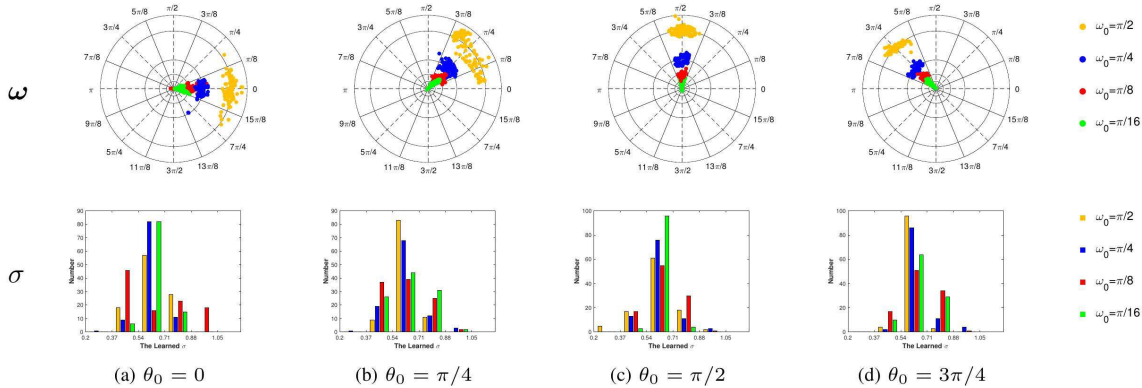}\\
\caption{Visualization of the learned frequencies (first row) and the histograms of the learned scales (second row) of the Gabor kernels in the first layer, where each column and each color show the learned parameters initialized with the same $\theta_0$s and $\omega_0$s, respectively, and all the $\sigma$s are initialized as $k/8$, \emph{i.e.}, $(5/8)$ in our experiments. That is, {the points and the bars marked with the same color in a column correspond to one of the 16 output features in the first layer.} The circles from inside out represent the cases where $\omega$ is $(\pi/16)$, $(\pi/8)$, $(\pi/4)$, $(\pi/2)$, and $2$, respectively. (Best viewed in color.)}
\label{fig:para:other}
\end{figure*}

\subsubsection{Parameters in Traditional Gabor Filters}

Here we analyze other parameters used in often-used hand-crafted Gabor filter construction, i.e, the frequency angle $\theta$, the frequency magnitude $\omega$, and the scale $\sigma$ in Gabor-Nets. Fig. \ref{fig:para:other} shows the learned angular frequencies determined by $\theta$s and $\omega$s, and the histograms of the learned $\sigma$s of the Gabor kernels in the first layer, where each color in each column corresponds to a kernel bank used to generate an output feature, i.e., $\mathbf{G}_{o}^{(1)}$. As shown in the first row of Fig. \ref{fig:para:other}, almost all the points gather in a $\theta_0$-centric sector region with an angle range of $(\pi/4)$, where those marked with different colors are well distributed around the arc corresponding to their $\omega_0$s.
Namely, although the points in Fig. \ref{fig:phase:freq} (b) tend to cover the whole semicircle region, the points representing different kernel banks barely overlap with each other. This means that, around  $\theta_0$ and  $\omega_0$, the Gabor kernel banks can extract the features with varying $\theta$s and $\omega$s rather than those intended for a single predetermined frequency (as the hand-crafted Gabor filters do), thus making Gabor filters in Gabor-Nets more powerful. Furthermore, as shown in the second row of Fig. \ref{fig:para:other}, the $\sigma$s are also automatically adjusted following the data characteristics during the learning process.

\section{Conclusions and Future Lines}  \label{sec:con}

We have introduced the naive Gabor Networks (or Gabor-Nets) for HSI classification which, for the first time in the literature, design and learn convolutional kernels strictly in the form of Gabor filters -- with much less parameters in comparison with regular CNNs, thus requiring a smaller training set and achieving faster convergence. By exploiting the kernel phase term, we develop an innovative phase-induced Gabor kernel, with which Gabor-Nets are capable to tune the convolutional kernels for data-driven frequency responses. Additionally, the newly developed phase-induced Gabor kernel is able to fulfill the traditional Gabor filtering in a real-valued manner, thus making it possible to directly and conveniently use Gabor kernels in a usual CNN thread. Another important aspect is that, since we only manipulate the way of kernel generation, Gabor-Nets can be easily implemented with other CNN tricks or structures. Our experiments on three real HSI datasets show that Gabor kernels can significantly improve the convergence speed and the performance of CNNs, particularly in scenarios with relatively limited training samples. However, the classification maps generated by Gabor-Nets tend to be over-smoothed sometimes, especially if ground objects are small and the interclass spectral variability is low. In the future, we will develop some edge-preservation strategies for Gabor-Nets to alleviate this negative effects. Furthermore, we will explore new kinds of filters that can be used as kernels in networks, which provides a plausible future research line for CNN-based HSI classification.

\bibliographystyle{IEEEtran} 
\bibliography{References}

\end{document}